\documentclass[useAMS,usenatbib, a4]{mn2e}

\usepackage{amsmath}

\usepackage{color}
\usepackage{graphicx}
\usepackage{hyperref}
\usepackage{booktabs}
\usepackage{float}
\usepackage{listings}

\def\beq{\begin{equation}}
\def\eeq{\end{equation}}
\def\mpch{$\text{h}^{-1} \text{Mpc}  $}
\def\hmpc{$\text{h} \, \text{Mpc}^{-1}  $}
\def\nh1{$N_{\text{HI}}  $}
\def\hi{H\,{\footnotesize I }}
\def\hii{H\,{\footnotesize II }}

\title [The effects of LLSs on EoR]
	{
		The Effects of Lyman-Limit Systems on the Evolution and Observability of the Epoch of Reionization
	} 

\author [H Shukla et al.]
	{
		Hemant Shukla,$^1$\thanks{\href{mailto:hshukla@lbl.gov} {email: hshukla@lbl.gov (affiliate)}}
		Garrelt Mellema,$^2$
		Ilian T. Iliev,$^1$
		Paul R. Shapiro$^3$\\		
		$^1$Astronomy Centre, Department of Physics \& Astronomy, Pevensey II Building,\\ 
			University of Sussex, Falmer, Brighton BL1 9QH, UK\\
		$^2$Department of Astronomy \& Oskar Klein Center for Cosmoparticle Physics, \\
			Stockholm University, Albanova, SE-10691 Stockholm, Sweden\\
		$^3$Department of Astronomy \& Texas Cosmology Center, \\
			University of Texas, Austin, TX 78712, USA
	}

\date{Released 2015 Xxxxx XX}

\pagerange{\pageref{firstpage}--\pageref{lastpage}} \pubyear{2016}

\begin{document}

\label{firstpage}

\maketitle

\begin{abstract} 

We present the first large-scale, full radiative transfer simulations of the reionization of the intergalactic medium in the presence of Lyman-limit systems (LLSs). To illustrate the impact of LLS opacity, possibly missed by previous simulations, we add either a uniform or spatially-varying hydrogen bound-free opacity. This opacity, implemented as the mean free path (mfp) of the ionizing photons, extrapolates the observed, post-reionization redshift dependence into the epoch of reionization. In qualitative agreement with previous studies, we find that at late times the presence of LLSs slows down the ionization fronts, and alters the size distribution of \hii regions. We quantitatively characterize the size distribution and morphological evolution of \hii regions and examine the effects of the LLSs on the redshifted 21-cm signal from the patchy reionization. The presence of LLSs extends the ionization history by $\Delta z \sim 0.8$. The LLS absorbers significantly impede the late-time growth of the \hii regions. The position dependent LLS distribution slows reionization further and additionally limits the late growth of the ionized regions. However, there is no ``freeze out" of the \hii regions and the largest regions grow to the size of the simulation volume. The 21-cm power spectra show that at large scales the power drops by a factor of 2 for 50\% and 75\% ionization stages (at $k = 0.1$ \hmpc) reflecting the limiting effect of the LLSs on the growth of ionized patches. The statistical observables such as the RMS of the brightness temperature fluctuations and the peak amplitudes of the 21-cm power spectra at large-scales ($k = 0.05 - 0.1$ \hmpc) are diminished by the presence of LLS.

\end{abstract}

\begin{keywords}

\hi, \hii regions: Lyman-limit Systems: High-redshift -- intergalactic medium -- cosmology, Epoch of Reionization: theory -- radiative transfer -- methods: numerical

\end{keywords}

%-----------------------------------------
% Section 1 - INTRODUCTION
%-----------------------------------------

\section{Introduction}

The first generations of baryonic structures, which formed before the
Universe was a billion years old, released sufficient numbers of hydrogen ionizing photons into the intergalactic medium (IGM) to completely ionize it. This process of cosmic reionization is generally assumed to be driven by ionizing photons produced by stellar sources, most of them with energies at or slightly above the Lyman limit. As a consequence, reionization was a very patchy process with ionized (H\,{\footnotesize II}) regions expanding around the positions of photon sources. The optical depth of the IGM initially defined the upper limit of the mean free path (mfp) of these ionizing photons to the distance at the edge of the ionized region within which their sources were located.

Over time, more sources formed creating newer ionized regions that grew and merged into an expanding patchy tapestry that eventually led to the completely ionized IGM at $z \sim 6$. In this post-reionization era the mfp for ionizing photons, measured directly from quasar spectra, is found to be shorter than it would have been in a uniform IGM with mean baryon density exposed to the average ionizing UV background - for a review on quasar spectra see \cite{Bechtold-2003}. The diminished value of mfp suggests the presence of discrete absorbing systems along the line-of-sight which have a high enough \hi column density to absorb most of the photons near the Lyman limit, i.e., $N_\mathrm{\hi} > 10^{17}$~cm$^{-2}$. These systems are known as Lyman Limit Systems (LLSs). The mfp due to these systems evolves with redshift and has been observationally determined over the redshift range $2.3 < z < 6$ \citep{Songaila-2010aa, Worseck-2014}.

Conceivably, similar optically thick systems should have been present during reionization. This implies that the LLSs would determine the mfp of the ionizing photons when the sizes of the ionized regions grow larger than the mfp. This, in turn, would result in a slowing down of the growth of ionized regions compared to the case where such systems are absent. Proper modeling of the reionization process should therefore include the effects of such LLSs. The mfp evolution for $z<6$ shows that the mfp at $z\sim 6$ is around 9 proper Mpc (pMpc) and rapidly declines with increasing redshift suggesting that the LLS will have a substantial impact on reionization.

The reionization process, however, is complicated by the fact that LLSs are not the sole absorbers of ionizing photons. During the EoR a significant fraction of photons is absorbed in the ionized IGM due to recombinations. As the density in the IGM fluctuates, so does the recombination rate and a proper estimate of the total effect of recombinations requires resolving the small scale density variations, including corrections for self-shielded systems (in which the highest density parts remain neutral and do not contribute to the recombinations) as well as the effect of photo-heating, which modifies the distribution of small scale density variations.

The distinction between LLS and recombinations in the ionized IGM is in fact somewhat artificial, as the lowest column density LLSs will be
high density but still ionized regions whose remaining \hi column densities give them optical depths larger than 1. At higher column
densities the systems will contain a fully neutral, high density core and can be described as self-shielding. Still, most of the absorption
in such systems happens in the ionized layer sitting around the neutral core. In a recent paper, \cite{Kaurov-2013} present an analysis of
the results of coupled radiation-hydrodynamic cosmological simulations in the context of ionized IGM density fluctuations and self-shielded systems in order to establish whether a distinction between their contributions as photon sinks during reionization has any physical basis. Although, they see a continuous variation in properties, they do conclude that such a distinction is physically motivated.

The small scale structures in the IGM cannot yet be resolved in the large volumes required for studying the reionization process ($>
100$~Mpc) and therefore need to be included using sub-grid recipes. This is far from trivial due to the complexities described
above and also because there are no observational measurements of the density and distribution of LLSs nor of the mfp for ionizing
photons during the EoR. As a result a range of different approaches has been used. These approaches can be divided into two main
categories, namely, (a) imposing a certain mfp for the ionizing photons, and (b) modifying the recombination rates to account for the unresolved density fluctuations. The former is equivalent to introducing discrete absorbers while the latter capture the effects of inhomogeneous recombinations throughout the IGM. 

The first category is a necessary ingredient for semi-numerical simulations \citep{Mesinger-2007, Choudhury-2009, Santos-2010, Mesinger-2010, Alvarez-2012} and has also been used in some numerical simulations \citep{Iliev-2014}. It is equivalent to imposing a hard limit on how far ionizing photons can travel but does not affect the photons for distances smaller than the mfp. Due to the uncertain evolution of the mfp for $z>6$ the value adopted is typically a free parameter, see \citet{Greig-2015}, which may or may not evolve with redshift.

Earlier numerical studies focused on so called mini-halos as discrete absorbers of ionizing photons \citep {Ciardi-2006, McQuinn-2007}, after detailed radiative-hydrodynamics simulations of mini-halo photoevaporation \citep{Shapiro-2004, Shapiro-2005} provided the evaporation times and photon consumption rates to assign to individual mini-halos overtaken by the global ionization fronts during the EoR. These were assumed to have a certain distribution and life time inside ionized regions and were then added as an additional
optical depth for the ionizing photons. The mfp values can be calculated from the assumed distribution of the absorbers. A related semi-analytical approach was taken in which this minihalo photon consumption was accounted for in the velocity of the I-fronts that lead the expansion of intergalactic \hii regions during reionization, including a statistical treatment of the biased clustering of minihalos surrounding the source halos \citep{Iliev-2005a, Shapiro-2006}. However, mini-halos will all have evaporated by the time reionization ends
\citep{Iliev-2005}. 

The second approach, i.e. accounting for unresolved recombinations through a sub-grid prescription, has been more commonly used in numerical simulations \citep{Iliev-2006, Iliev-2007, Iliev-2008, Kohler-2007} and was recently introduced in semi-numerical simulations by \cite{Sobacchi-2014}. In this approach the mfp is not imposed but has to be calculated from
the simulation results. 

Although the details vary, all of the above studies find that the size distribution of ionized regions is affected by the presence of the LLS. Especially at the later stages of reionization, the ionized regions do not grow as fast as when no LLS are included. This typically leads to a reduction of the large scale power in power spectra.

Very few large scale numerical reionization simulations have included discrete absorbers, the exceptions being the ones mentioned above. This means that a number of questions remain unanswered. In this paper we want to address some of these questions using an improved algorithm to limit the mfp of ionizing photons. The questions we focus on are,

\begin{itemize}
\item [(i)] How a redshift dependent mfp, as suggested by the post-reionization
observations, affects the reionization process? We use the homogeneous distribution of the LLS for examining the effects.

\item [(ii)] Whether a density dependent distribution of absorbers changes their impact? We implement the density (position) dependent LLS to explore the same.
\end{itemize}

The units in this paper are defined such that all the lengths in the units of \mpch \ or \hmpc \ are comoving unless mentioned otherwise. The cMpc and pMpc are comoving and proper Mpc respectively.

This paper is organized in the following sections. In \S~\ref{sec-2} we discuss the Lyman-limit systems and their observations. The N-body and radiative transfer simulations along with the Lyman-limit systems implementation are presented in \S~\ref{sec-3}. The size distribution of the \hii regions and the associated power spectra derived from our simulations are discussed in \S~\ref{sec-4}. In \S~\ref{sec-5} we present the effects of LLSs on the 21-cm observables. We compare the results of the position dependent implementation of the LLS in~\S~\ref{sec-6} followed by our conclusions in~\S~\ref{sec-7}. 

%-----------------------------------------
% Section 2 - LLS
%-----------------------------------------

\section{Lyman-limit Systems} \label{sec-2}

The \hi column densities, estimated by the absorption lines of the intervening hydrogen observed in post-reionization quasar spectra, are used to categorize the hydrogen absorption systems into three overlapping states, \emph {viz.}, Lyman-$\alpha$ forest (for \hi column densities \nh1 $< 10^{17} \rm cm^{-2}$), Lyman-limit systems (LLSs, $10^{17} \rm cm^{-2} <$ \nh1 $< 10^{20} \rm cm^{-2}$), and damped Ly-$\alpha$ systems (DLAs, \nh1 $>10^{20} \rm cm^{-2}$). The Ly-$\alpha$ forest consists of low density and highly ionized structures, in contrast to DLAs, which are high density and partly neutral, thus exhibiting a strong damping wing of the Ly-$\alpha$ line. The studies of these systems have enabled precise measurements of the \nh1 values leading to high precision constraining of observables such as the primordial power spectrum. 

The Lyman-$\alpha$ forest and DLAs do not affect the reionization process significantly due to their low optical depth (forest) and relative rarity (DLAs). In contrast, LLSs have both a relatively high optical depth and abundance, and thus the potential to considerably influence the later stages of the reionization \citep{Alvarez-2012}.

First observed as quasar absorption lines in surveys~\citep{Tytler-1982}, the Lyman-limit systems appear as abrupt discontinuities in the quasar absorption line spectra at the rest-frame Lyman limit at wavelength $\lambda\sim$ 912 \AA. \cite{Prochaska-2010} define LLS as regions with Lyman continuum optical depth of $\tau_\text{LLS} \geq 2$, \emph{i.e.}, \nh1 $\geq 10^{17.5}$ cm$^{-2}$.   The LLSs are assumed to be located in and around galactic halos. These systems are relatively easily identifiable even with low resolution and poor signal-to-noise. However, unlike the Ly$\alpha$ forest and DLAs, the LLSs are poorly understood largely because \nh1 estimations require complete spectral coverage of both the Ly$\alpha$ line and the Lyman break. At lower redshifts ($z < 2.6$) the Lyman limit is shifted into the UV spectrum and thus is unobservable from the ground. High redshift surveys~\citep{Songaila-2010aa, Prochaska-2010} present measurements of the number density function of the LLS. Another recent survey~\citep{Ribaudo-2011} with Hubble Space Telescope archival data identifies 206 LLSs for $z < 2.6$.

All these surveys identify the LLSs in the absorption line spectra and estimate the number of LLSs per unit redshift per \hi column (\nh1) density function - $f (N_\text{HI}, z) \propto \partial^2 \mathcal{N} / \partial z \partial N_\text{HI}$. The best fit power-law index, $\beta$, to this function constraints the column densities of the LLSs. In addition, earlier  simulation studies~\citep{Kohler-2007, McQuinn-2011aa, Sobacchi-2014} of LLS abundances and the mfp of ionizing photons agree reasonably well with the observations.
Using the parametric values of the density function from the observations~\citep{Songaila-2010aa} and a simulation model~\citep{McQuinn-2011aa} we define two simulation models, LLS1 and LLS2 respectively (see \S~\ref{lls:sect} and Table~\ref{table:lls} for details). 

\begin{figure}
\centering \includegraphics[scale=0.42]{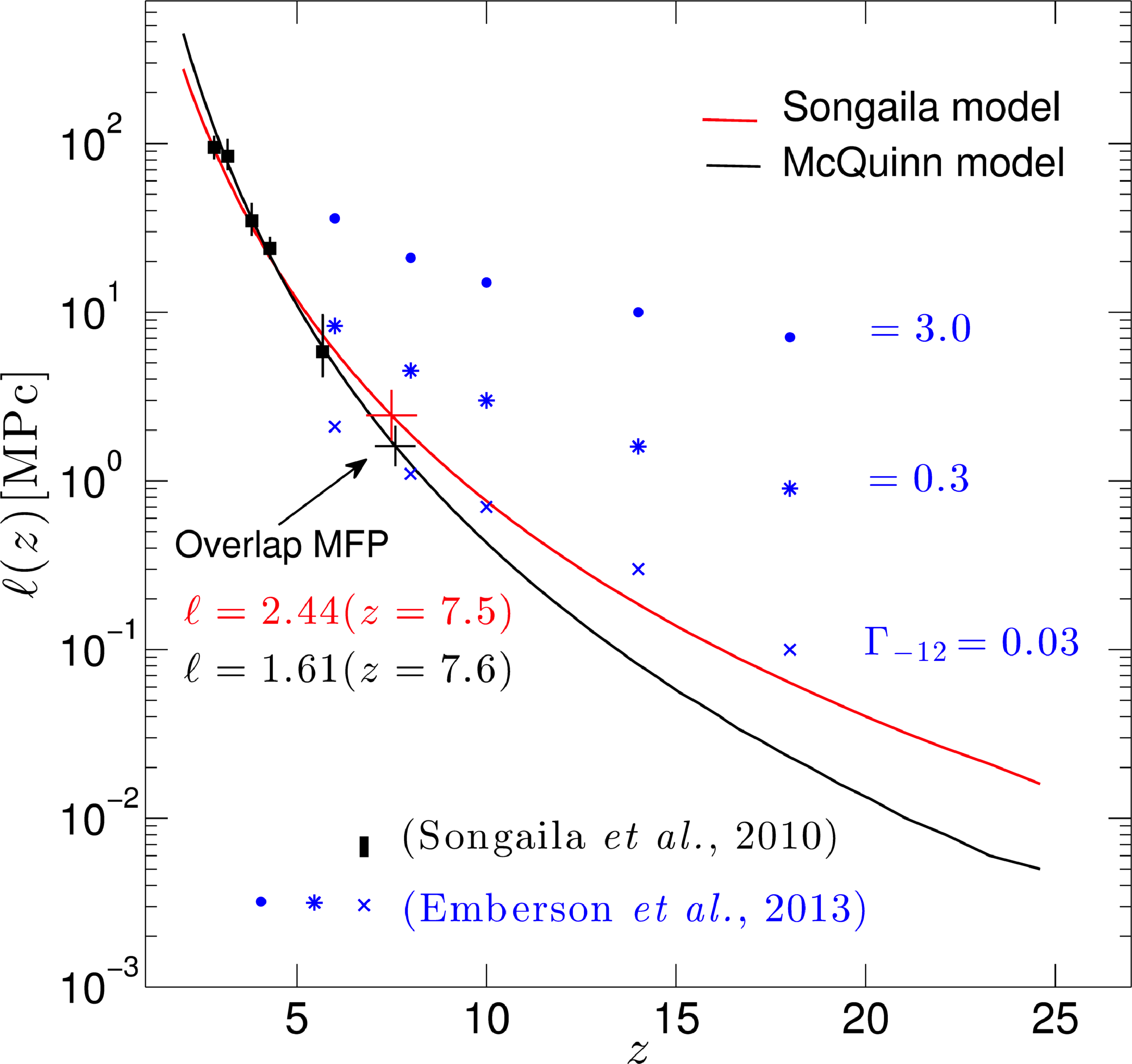}
\caption{The mean free path (in log$_{10}$ scale) of the ionizing photons for the two LLS models in solid lines, Songaila (red), and McQuinn (black) in proper Mpc. The open circles show the mfp at the overlap for the simulations (LLS1 and LLS2) based on the two models. The simulation parameters are listed in Table~\ref{table:lls}. The filled black squares are the data points from \citet{Songaila-2010aa}. The blue symbols are from the ionization models from \citet{Emberson-2013} for three values of the ultraviolet background ($\Gamma_{-12}$). }
\label{mfp}
\end{figure}

The general effect of the LLSs is to restrict the mfp of ionizing photons and consequently impede the evolution and merging of \hii regions during reionization.  Early on during reionization, when the ionized regions are still small, the mfp is determined by the size of these regions. Towards the end of the reionization, the \hii regions reach sizes larger than the measured mfp due to LLS at $z\sim6$, of approximately 60 comoving Mpc and thus the mfp should be set by the LLS. This is the stage where the LLSs begin to regulate the ionization history. At late times, there are many groups of local ionizing sources that add to the ionization fronts further complicating the morphological evolution of the ionized regions. The simulations discussed herein attempt to quantify the effects of the LLSs on the reionization history, with the caveat that the lack of high redshift observational data leads to an ad hoc implementation of the LLSs at early times. The extrapolation of mfp from observed low redshifts to early times as far as $ z = 20$ clearly yields unrealistic values, see Figure \ref{mfp}. However, as discussed below, LLSs only start to affect the simulations much later after $z = 15.96$ (LLS1) and $z = 13.30$ (LLS2) with mfp value of 0.1 Mpc in proper units. More accurate effects of LLS at higher redshifts could only be modeled based upon the actual distribution of the high redshift LLSs. However, the very first numerical simulations of a large volume and higher dynamic range presented here, elucidates various useful insights beneficial for future research in the field. 

%-----------------------------------------
% Section 3 - SIMULATIONS
%-----------------------------------------

\section {Simulations} \label{sec-3}
The evolving matter density fields in a given comoving volume for the desired redshift ranges are generated with the N-body code entitled CubeP$^{3}$M~\citep{Deraps-2013}. The evolution of the density fields is based upon the initial conditions specified by the standard Zel'dovich approximation and primordial power spectrum transfer function derived by CAMB \footnote{http://camb.info} \citep{Lewis-1999}, originally based on CMBFAST~\citep{Seljak-1996}. The cosmological parameters used are for the flat $\Lambda$CDM model of the Universe based on WMAP 5-year data combined with constraints from baryonic acoustic oscillations and high-redshift supernovae, given as, $(\Omega_\text{M} = 0.27, \Omega_{\Lambda} = 0.73, h = 0.7, \Omega_b = 0.044, \sigma_8 = 0.8, n_s = 0.96)$. To ensure against numerical artifacts~\citep{Crocce-2006aa} the initial conditions are generated at sufficiently high redshift (here $z_i = 300$).
\begin{table}
\begin{center}
\begin{tabular}{c c c c c c} \\ \toprule
 box size 		&  $N_\text{part}$   	& mesh     &  spatial  	& particle  & min($M_\text{halo}$)  \\ 
 & & & resolution & mass & \\
 \mpch& & & h$^{-1}$kpc & $10^6 M_{\odot}$ & $10^8 M_{\odot}$ \\
$114$	&  $3072^3$		& $6144^3$      & $1.86$	& $5.47$ 	&  $1.09$\\ 
\bottomrule
\end{tabular}
\end{center}
\caption{ The N-body simulation parameters for the two volumes.}
\label{table:n-body}
\end{table}

The CubeP$^{3}$M code is public domain N-body code designed for simulating large-scale cosmological systems. The code is accurate, efficient, scalable, and parallel across distributed (MPI) and shared (OpenMP) memory systems. The underlying N-body algorithm estimates short-range (sub-grid distances) gravitational forces using the particle-particle (P-P) method. While for the long-range forces, a 2-level particle-mesh (PM) method is applied. Computationally, in comparison to the P-P method which is of the order $O(N^2)$, the P$^3$M method has significantly lower overhead and is of the order $O(N\log{N})$, where $N$ is the number of particles.

For the current study, two sets of comoving volumes of sizes, 37 \mpch\  and 114 \mpch, are used. The same N-body simulations were presented in detail in \citep{Iliev-2012}. The smaller volume, 37 \mpch, is used for testing and validation purposes only. The results from the smaller volume are not included in this paper. 

The simulations for 114 \mpch \ use $3072^3 \approx 28.9$ billion dark matter particles distributed in $6144^3$ mesh cells. 
Each particle has mass of $5 \times 10^6 M_{\odot}$. Through the simulation steps as the structures start to form, the halos are identified using a spherical overdensity halo finder with overdensity parameter of $\Delta=178$ with respect to the mean density. The halos with more than 20 particles ($M > 10^8 M_{\odot}$) are considered resolved. The number density of the halos increases for lower redshifts and the mass function approaches the Sheth-Tormen mass function. \cite {Iliev-2006aa} and \cite {Watson-2013} provide detailed fits to the high-redshift  halo mass function.

%% Figure

\subsection {C$^2$-Ray - Radiative Transfer}
\label{sec:c2ray}
The second stage of the simulation performs radiative transfer using the $\operatorname{C^2-Ray}$ (Conservative, Causal Ray-Tracing) code~\citep{Mellema-2006a}. The conservative part of the code ensures spatial and temporal photon conservation, while the causal ray-tracing is implemented using the short-characteristic method. 

The C$^2$-Ray implements a discrete spatial and temporal
version of the ionization rate equation as \citep{Osterbock-1989}, 
\begin{equation}
\Gamma(r) = \frac {1}{4\pi r^2} \int_\nu {\frac{L(\nu) \sigma(\nu) e^{-\tau(\nu,r)}} {h\nu} \ d\nu}
\label{gamma1}
\end{equation}

where, $\Gamma (r)$ is the ionization rate at distance $r$ from the
hydrogen ionizing source, $L_\nu$ is the spectral energy distribution
of the ionizing source at frequency $\nu$, $\sigma_\nu$ is the
cross-section for the ionizing photons, and $\tau_{\nu}$ is the
frequency dependent optical depth of the hydrogen gas for bound-free transitions.

Due to the spatial discretization of the computational domain, we do not
know $\tau_{\nu}$ as a continuous function of position. As was shown in
\citet{Abel-1999}, the photo-ionization rate of one cell whose center has a distance $r$
from the source can be calculated as,

\begin{equation}
  \Gamma = \frac {\dot{N}(r - \frac {\Delta r}{2}) - 
    \dot{N}(r + \frac {\Delta r}{2})}{n_\text{H\,I} V_\text{shell}}
\label{gamma2}
\end{equation}

where, $V_\text{shell}$ is the volume of the spherical shell with radius $r$ and width $\Delta r$, the shell is filled with neutral hydrogen of number density  $n_\mathrm{H I}$, $\dot{N}(r - {\Delta r}/{2})$ is the rate of ionizing photons arriving and $\dot{N}(r + {\Delta r}/{2})$ the number of photons leaving this
shell. 

Since,

\begin{equation}
  \dot{N}(r) =  \frac {1}{4\pi r^2} \int_\nu {\frac{L(\nu) e^{-\tau(\nu,r)}} {h\nu} \ d\nu}\,,
\end{equation}

Equation~\ref{gamma2} implies that the local photo-ionization rate $\Gamma$ depends
on the difference between $\tau_\mathrm{in}\equiv\tau(r - \frac {\Delta r}{2})$
and $\tau_\mathrm{out}\equiv\tau(r + \frac {\Delta r}{2})$. In Section \ref{sect:algol} we
explain how these optical depths are modified to include the effect of LLS.

In our reionization model, the ionizing luminosity of the collapsed halos is
proportional to their mass $M$. Each halo produces a number of photons,

\begin{equation}
  N_\gamma = \frac{f_\gamma M \Omega_b}{\Omega_0 m_p}
\label{N_gamma}
\end{equation}

for every n-body output of $\Delta t=11.46$~Myr. The efficiency factor
$f_\gamma$ is the product $f_{esc} f_{\star} N_{\star}$, where,
$f_{esc}$ is the ionizing photon escape fraction, $f_\star$ is the
star formation efficiency, and $N_\star$ is the number of ionizing
photon per stellar atoms, and $m_p$ is the proton mass. The parameter $N_\star$
depends on the initial mass function (IMF) of the stellar population
producing the ionizing radiation. Its value for a Pop II population
(Salpeter IMF) is $\sim$4000 and for a Pop III population (top-heavy
IMF) it can reach $\sim$100,000. Due to the uncertainties in $f_{esc}$
and $f_{\star}$, the value of $f_\gamma$ is not well
constrained. In this study we use 10 for high mass halos (HMACH: High
Mass Atomicaly-Cooling Halos) and 150 for low mass halos (LMACH:
Low Mass Atomicaly-Cooling Halos), see Table~\ref{table:reion}.
The higher value for the LMACHs is motivated either by a larger
contribution of metal-free/poor stars or by a larger escape fraction.
The LMACHs are also assumed to be susceptible to negative radiative 
feedback. When the cell in which an LMACH is present is ionized at the
start of a new 11.46 Myr time step, the LMACH will not produce any
ionizing photons.

From previous work we know that the efficiency factors chosen result
in reasonable reionization history in accord with the WMAP optical
depth value of $\tau_\mathrm{es}=0.089 \pm 0.014$
  \citep{Hinshaw-2013}. We note that the recent results from the
  Planck mission favour a lower value of $0.066 \pm 0.016$
  \citep{Ade-2015-b} for this parameter.

\begin{table}
\centering
\begin{tabular}{c c c c } \\ \hline
 box size 	&  $f_\gamma$ & $f_\gamma$ &  RT grid \\
 \mpch	&  HMACH	       & LMACH         & \\ \hline
114		&  10		       & 150               & $256^3$\\ \hline
\end{tabular}
\caption{ Simulation parameters for the 114 \mpch\  box with LLS. $f_\gamma$ is the star formation efficiency for high and low mass, and RT grid is the coarser grid for radiative transfer ray-tracing. The underlying cosmology uses the WMAP 5-year results.}
\label{table:reion}
\end{table}

\subsection {Simulating the Effects of Lyman Limit Systems}
\label{lls:sect}

It is thought that the LLSs correspond to the denser, ionized gas associated with collapsed objects and dense filaments of the cosmic web which, although ionized, have sufficiently high column density of \hi to result in an optical depth $> 1$. This optical depth is usually described in terms of a mean free path for the ionizing photons and the available observational constraints are expressed in the same way. 

There are two possible approaches to implementing the effects of the LLSs in our calculations - either as an enhanced local recombination rate, or as an additional absorber optical depth along each light ray. The first approach is closer to the physical mechanism responsible for these objects, and recombinations in dense structures below the resolution limit of the code are described with clumping factors. This has for example been the approach of \cite{Sobacchi-2014}. Introducing a local clumping factor to reflect an enhanced recombination rate does not however directly relate to a value of the mean free path (mfp) due to such absorbers. It is thus difficult to impose such a constraint on the LLS models used in the simulations. Therefore, for this study we decided to work directly with the mfp instead. This approach is similar to prior (semi-numerical) work by \cite{Alvarez-2012}, however, these authors implemented the mfp as a hard boundary which photons could not cross, which is not very physical. In our implementation the mfp defines an additional ionizing photon-absorbing component which after one mfp reaches an optical depth of 1. This approach allows us to directly use the published observational mfp expressions. An additional advantage of working with the mfp directly is that it is easier to connect the effect on the sizes of \hii regions to the imposed mfp.

The mfp in the ionized medium is not a measured quantity beyond $z=6$, however. Therefore, we extrapolate it from the lower redshift results. To the extent that the ionized regions during reionization can be viewed as being locally post-reionization, there is some justification in performing this extrapolation. The additional absorption that the implemented LLS component adds in the fully neutral regions is marginal, and therefore only affects the already ionized regions. To perform the extrapolation of the mfp beyond $z=6$, we use the parametrization given by \cite{Songaila-2010aa}.  Based upon the observational data, the number density of the LLSs per unit redshift path $\text{d}z$ is parametrized as,

\begin{equation}
f(N_\text{HI}, z) = f(N_\text{HI}, z=3.5) \left( \frac {1+z} {4.5} \right) ^\gamma
\end{equation}

where, $f(N_\text{HI}, z = 3.5)$ is the number density at $z = 3.5$. Estimating the log-likelihood function for the entire redshift range of $0 < z < 6$ in~\citet{Songaila-2010aa} yields the values for the parameters to be $ f(N_\text{HI}, z = 3.5) = 2.8 \pm 0.33$ and $\gamma = 2.04_{-0.37}^{+0.29}$. Furthermore, in the approximation that the column density function $f(N_\text{HI}, z) dN_\text{HI} \propto N_\text{HI}^{-\beta} dN_\text{HI}$~\citep{Petitjean:1993}, the mean free path is related to the number density as~\citep{Miralda-Escude-2003},

\begin{align}
\ell(\nu_0, z) f(N_\text{HI}, z) &\propto \frac {\int_a^\infty \tau^{-\beta} d\tau}{\int_0^\infty \tau^{-\beta} (1 - e^{-\beta}){d\tau}}\\
\text{or}, \quad
\ell(\nu_0, z) &=  \frac{a^{1-\beta}}{\Gamma(2-\beta)} \ \frac{c}{H(z)(1+z)f(N_\text{HI}, z)}
\label{eqn:ell}
\end{align}

where, $f(N_\text{HI}, z)$ is measured above the column density $N_\mathrm{HI}$ corresponding to a value $a$ for the optical depth (here, $a \approx 1$), $\tau$ is the optical depth at the Lyman limit,  and $\nu_0$ is ionization edge frequency. Here, $\Gamma$ is the gamma function and not the ionization rate~\citep{Songaila-2010aa}.

For simulation L1 (without LLS) at z = 8.39 (overlap), the average Lyman limit optical depth is 0.83 over the volume size of 114 \mpch. This translates into an mfp of 20.6 pMpc. This value is an order of magnitude larger than expected from an extrapolation of the observed lower redshift values. This is due the fact that our radiative transfer resolution of 0.45 \mpch \ does not resolve the sub-grid density inhomogeneities which determine the actual mfp value.

In order to test the impact of different evolutions of the mfp we use two different extrapolations, one using the parameters given by \cite{Songaila-2010aa} and one using parameters derived from fitting the curve in the inset of Figure~1 in \cite{McQuinn-2011aa}. The simulations for these two choices are labeled as LLS1 and LLS2, respectively and the parameters used are listed in Table 3. The evolution of the mfp for these two sets of parameters is shown in Figure 1. LLS2 has substantially smaller values of the mfp through the reionization
epoch. The figure also shows the values of the mfp at different redshifts below 6 as estimated by \cite{Songaila-2010aa} 
(using $\beta=1.28$ and $\gamma=1.94$ in their equation 7).

For comparison, we also show the evolution of the mfps for three different photoionization rates in terms of $\Gamma_{-12}$ - measured $\sim 0.3$ s$^{-1}$ for Ly$\alpha$ forest at $z\sim6$ - from the simulations of \citet{Emberson-2013}. Note that these mfps are based on the Lyman limit optical depth from the radiative transfer calculations of ionization by a fixed radiation background in a 0.5 Mpc sized box.

From Figure 1 it is obvious that both extrapolations result in very small mfps at very high redshifts. Below our cell size it does not
make sense to implement the LLS model as we already use an assumed escape fraction for absorptions within the source cell. Furthermore,  at very high redshifts the cosmic structures are much less developed, thus fewer LLS systems should exist, implying longer mean free paths due to these systems.  It is likely, therefore, that these extrapolations become unreliable at  such high redshifts. As we want to concentrate on the later stages of reionization when the ionized regions have reached sizes of 10-20 cMpc  and LLSs should have more impact, we chose to only switch on our LLS absorption if the mfp is larger than 5 grid cells (3~cMpc). This occurs  around $z\sim 15$. We found that this choice did not impact the evolution around this transition redshift.

\begin{table}
\centering
\begin{tabular}{l c c c c c c} \\ \hline
 Simulation & Model	&  $\gamma$ & $f (N_\text{HI}, z_\text{x}) $ &  $z_\text{x}$ & $\beta$ \\ \hline
LLS1 & Songaila \emph{et al.}  & 2.04 & 2.84 & 3.5 & 1.28 \\
LLS2 & McQuinn \emph{et al.}  & 2.85 & 2.34 & 3.5 & 1.30\\ 
L1 & No LLS & - & - & - & -\\ \hline
\end{tabular}
\caption{ Simulation parameters for the two LLS simulations LLS1 and LLS2. In simulation L1 there are no LLS added.}
\label{table:lls}
\end{table}
 
\subsection {LLS implementation in C$^2$-Ray}
\label{sect:algol}

To include the effects of LLS due to mfp in the radiative transfer calculation, an
additional optical depth term, $\tau_\mathrm{LLS}$, is
added. The implementation of the additional optical depth is based upon the notion that after one mean free path, $\tau_\mathrm{LLS}$
acquires the typical value of the optical depth of a Lyman-limit
system at the Lyman limit. The opacity may be added in two different ways depending upon the assumptions made for the spatial distribution of LLSs. In the first scenario, a uniform distribution of the LLSs is assumed, while in the second,  the LLSs are assumed to be concentrated in the higher density regions. The latter case is motivated by the observation that LLSs are likely to be located at the outer regions of the halos, embedded in the filaments of the large scale structure \citep{Furlanetto-2006, Erkal-2014}.

In the uniform distribution of LLSs, we assume
that each cell contributes equally to $\tau_\mathrm{LLS}$ with a value
$\Delta \tau_\mathrm{LLS}$. In C$^2$-Ray there are two optical
depth values associated with each cell. The value, $\tau_\mathrm{in}$, is the optical depth
between the source and the entry point of the ray into the cell, and
$\Delta \tau$, is the optical depth of the ray that is contained
within the cell. The difference between these two values is used to
calculate the ionization rate $\Gamma$ in the cell, see Equation~\ref{gamma2}. The sum of the two values is the input optical depth of the adjacent cell along the line of traversal of the ray.

For two cells, $n$ and $n+1$, with the same ray crossing their centers, the input optical depth for cell $n+1$,
$\tau_\mathrm{in}^{n+1}$, is equal to the output optical depth of cell $n$, $\tau_\mathrm{out}^n \equiv \tau_\mathrm{in}^{n} +
\Delta \tau^n $. To include the LLS optical depth this equality is
changed to
\begin{equation}
  \tau_\mathrm{in}^{n+1} = \tau_\mathrm{out}^{n} + \Delta \tau_\mathrm{LLS}\,.
\end{equation}
This effectively adds the additional optical depth due to LLS in between the two
cells which means that the LLS optical depth is not used in the calculation of $\Gamma$
for a cell. In principle, one could add $\Delta \tau_\mathrm{LLS}$ to $\Delta \tau$
but this would then impact $\Gamma$ which would give incorrect
results. The fundamental purpose to implement the sub-grid LLS model is to remove (absorb)
photons due to LLS without impacting any other parts of the radiative transfer
calculation.

For the position dependent implementation of the LLS in C$^2$-Ray, the primary difference is in the estimation of the $\Delta \tau_\text{LLS}$. The value of $\Delta \tau_\text{LLS}$ is proportional to the normalized sum of the cross sections of all halos in a cell. The normalization is such that the average of all the cells has the average total geometric cross-section required to reproduce the assumed mean LLS opacity per cell. The cross section is defined as $\pi r_\text{vir}^2$, where $r_\text{vir}$ is the virial radius for a halo. The opacity in a cell is proportional to the cross section and the mean LLS opacity (over all cells), which is set by the fit to the redshift evolution of the mfp.
The assumption is that larger halos with larger virial radii have more in terms of LLS structures associated with them. All halos are used for this estimation. Even though ionization may suppress star formation in the low mass halos, presumably the density fluctuations will survive better. 

Given the assumed redshift evolution of the mfp of LLS {$\ell(\nu_0,z)$, see Equation~\ref{eqn:ell}, we estimate the average additional optical depth per cell as,
\begin{equation}
  \Delta \tau_\mathrm{LLS} = \tau_\mathrm{LLS}\frac{\Delta x}{\ell(\nu_0,z)}\,.
  \label{Delta_tau}
\end{equation}

where, $\Delta x$ is the cell size. For the homogeneous case, this is the additional optical depth per cell. For the position dependent case, this is the average value per cell, with the actual value for a cell being proportional to the total cross section of halos in that cell. the cells with a larger total cross section for halos getting higher values and the cells without any halos getting none. Equation~\ref{Delta_tau} ensures that the after a distance $\ell(\nu_0,z)$ all rays will have picked up an additional optical $\tau_\mathrm{LLS}$. We take  $\tau_\mathrm{LLS}$ = 2.

For the cells whose centers do not lie on the same ray, the short characteristic ray tracing algorithm constructs $\tau_\mathrm{in}$
through interpolation of the relevant $\tau_\mathrm{out}$. The above modification works equally well if one replaces each $\tau_\mathrm{out}$ by $\tau_\mathrm{out}^{n} + \Delta \tau_\mathrm{LLS}$.

As explained in Section \ref{lls:sect}, this additional optical depth is only applied when $\ell(\nu_0,z) > 3$~cMpc. For smaller values we set $\Delta \tau_\mathrm{LLS}$ to zero.

In the following sections we use the fiducial homogeneous distribution implementation of the LLSs for various studies. The comparison of the homogeneous and position dependent methods is discussed in~\S~\ref{sec-6}.

%-----------------------------------------
% Section 4 - RESULTS
%-----------------------------------------
\section {Results} \label {sec-4}

In this section we summarize the results of the various analysis methods we use to characterize the simulations and compare them between the L1 (no LLS), LLS1 and LLS2 cases.

%%Figure
\begin{figure}
\centering \includegraphics [scale=0.36]{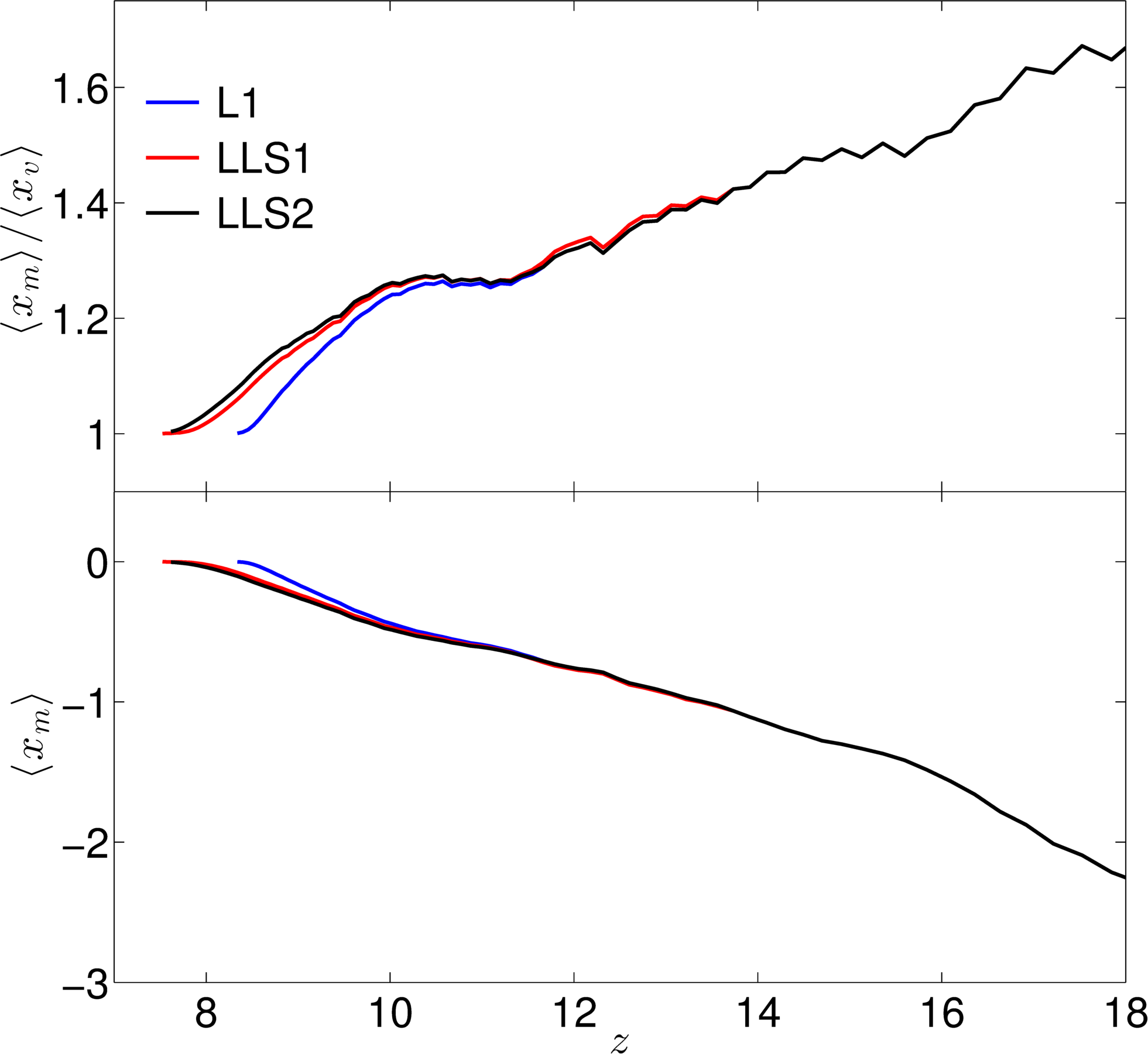}
\caption{ (Top) The ratios of mass- and volume-weighted, $x_v$, vs. redshift $z$; and (Bottom) Evolution of the ionized fractions: log of mass-weighted, $x_m$, for the three simulations L1, LLS1, and LLS2.}
\label{xmxv}
\end{figure}

\subsection{Globally Averaged Quantities} \label {ssec-41}

Figure~\ref{xmxv} shows the evolution of the globally averaged mass-weighted $\langle x_m \rangle$ and volume-weighted $\langle x_v \rangle$ ionized fractions of the three models as a function of redshift. In both the top and the bottom panels of the figure we note that the models start diverging at $z = 14$ and that the differences become more pronounced after $z \sim 10-11$. The shorter mean free paths imposed by the LLSs delay the overall process of the expansion and merging of the \hii regions.  The simulation L1 (without LLS) reaches a global average volume-weighted ionized fraction of $\langle x_v \rangle = 0.98$ at $z = 8.34$, while for LLS1 and LLS2, the same level of ionization is reached at $z = 7.61$, and $z = 7.71$, respectively. The top panel shows the ratio $\langle x_m \rangle/\langle x_v \rangle$ while the bottom panel shows only the averaged $x_m$. From the top panel it is evident that the mass-weighted ionized fraction is always significantly higher than the volume-weighted fraction for all the three models. This means that reionization is inside-out, that is, the dense regions surrounding the sources are preferentially ionized first yielding higher $x_m$ averages. The ionization fronts expand outwards, eventually reaching the less dense regions and voids. The ratio $x_m/x_v$ is the mean over-density of the ionized regions \citep{Iliev-2006aa} as shown below,

\beq
\frac{x_m}{x_v} = \frac{V_\text{box}}{M_\text{box}} \ \frac{x_m M_\text{box}}{x_v V_\text{box}} = \frac{1}{\bar{\rho}} \frac{M_\text{ionized}}{V_\text{ionized}}
\eeq

where, $\bar{\rho}$ is the mean density of the Universe. Figure~\ref{xmxv} indicates that the over-density of the ionized regions is always larger than one and for the LLS cases is even larger than for L1. This is because less dense regions such as voids do not have ionizing sources and therefore require photons from afar to get ionized. The simulations LLS1 and LLS2 reach a numerical value of 1.0 at redshifts $z = 7.76$ and $z = 7.617$ respectively, while for the L1 model this value is reached much earlier at redshift $z = 8.397$, thereby, delaying reionization by $\Delta z = 0.78 - 0.64$. 

%% Figure
\begin{figure}
\centering \includegraphics[scale=0.47]{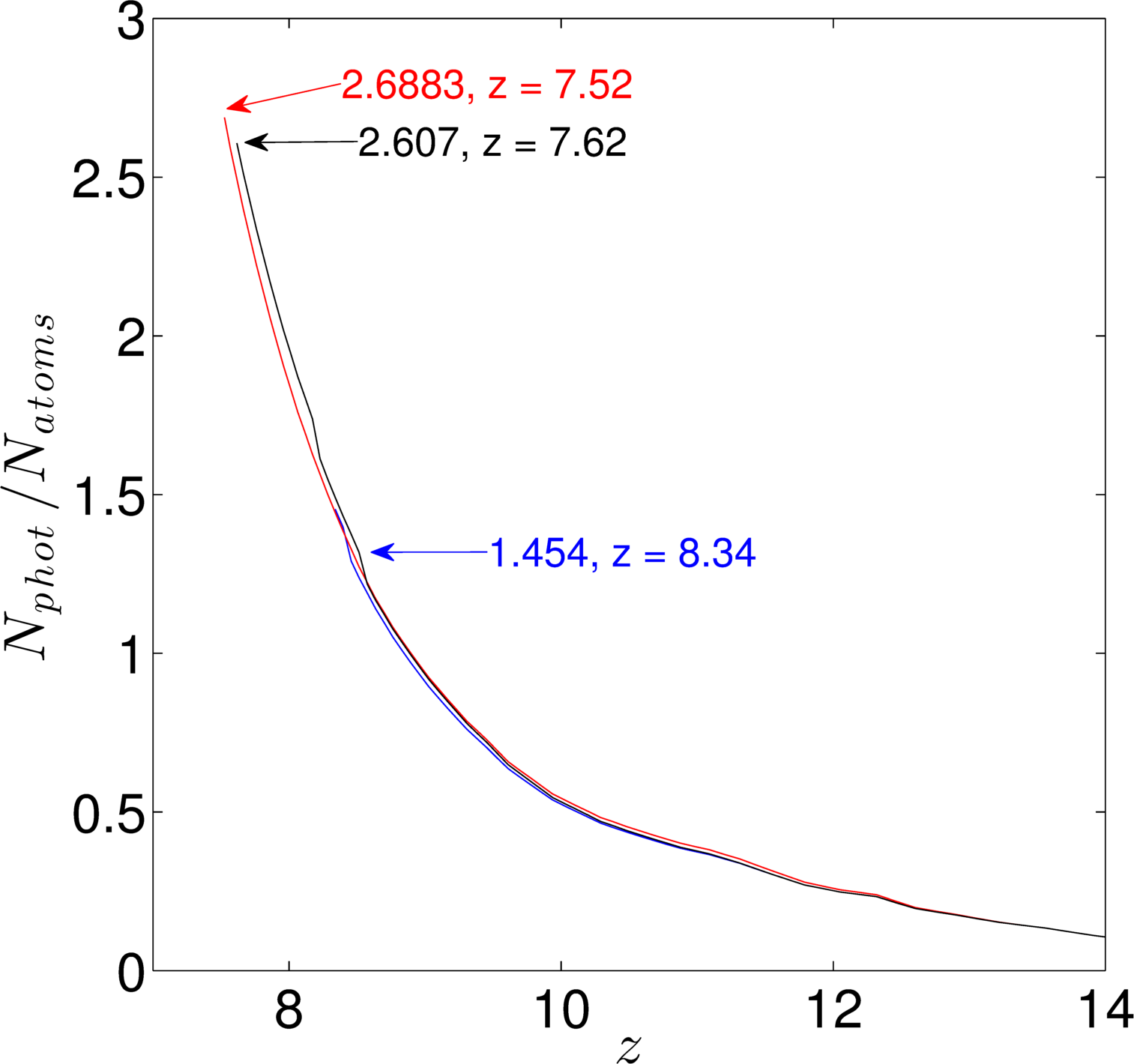}
\caption{The cumulative number of ionizing photons per total gas atoms in the simulation volume for the three cases L1 (blue), LLS1 (red), and LLS2 (black). The arrows label the respective $N_{phot} /N_{atoms}$ values and the corresponding redshifts when reionization completes.}
\label{phot_atoms}
\end{figure}
\subsection {Photon Statistics}

%%Figure
\begin{figure*}
\centering \includegraphics[width=6.2in]{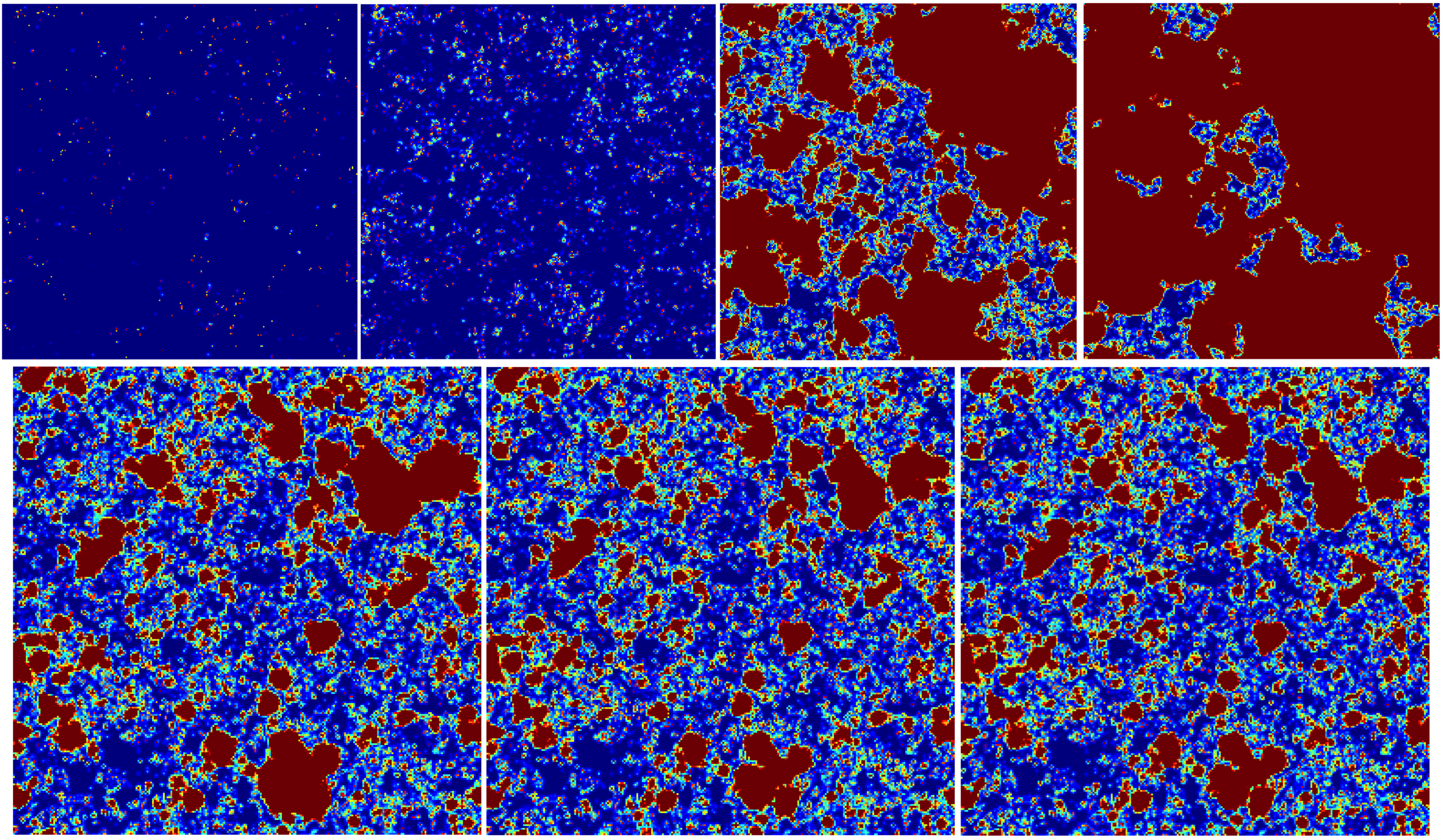}
\caption{(Top) Spatial slices of the ionized hydrogen for the total 114 \mpch \ box for the L1 case (no LLSs) at different ionization fractions and redshifts. From left to right - $\langle x_m \rangle$ = 0.1, 10.0, 75, and 95\% at $z = $ 16.9, 13.2, 8.892, and 8.515. (Bottom) Spatial slices for three models at $\langle x_m \rangle$ = 50\%. From left to right, models L1, LLS1, and LLS2 at redshifts $z$ = 9.4, 9.3, and 9.1.}
\label{xfrac-all}
\end{figure*}

The reionization period is defined by the complex interaction and evolution of the ionizing photon sources and sinks. In the simulations the sources of the ionizing photons are the halos. The very first HMACHs form around $z \sim 21$. For a more in-depth study of the mass and the number density distribution of the halos, see \citet{Iliev-2012}. The clustering of halos defines the photon emanating regions and sets the initial conditions for the formation and evolution of \hii regions. Initially, the mean free path is dominated by the size of the ionized regions in the IGM. While the LLSs continuously absorb ionizing photons, it is only at later times ($z \sim 14-10$) that the absorption contributions due to the LLSs become significant.
 
The photon and baryon populations in the simulations are recorded to extract statistical properties of interest. The integrated Thompson electron-scattering optical depth $\tau_\text{es}$ for the three cases as a function of redshift as compared to WMAP-9 mean optical depth estimates lie within the 1-$\sigma$ spread. They also fall within 1-$\sigma$ range of the Planck estimate ($0.066 \pm0.016$). The values of the optical depth, $\tau_\text{es}$, in the simulations at $z = 25.33$ for L1, LLS1, and LLS2 are 0.0819, 0.0796, and 0.0788, respectively. The optical depth in the presence of LLS is thus diminished by about $\sim 0.002$. This is expected as there are overall less ionized electrons available for scattering the CMB photons.

%%Figure
\begin{figure}
\centering \includegraphics [scale=0.46]{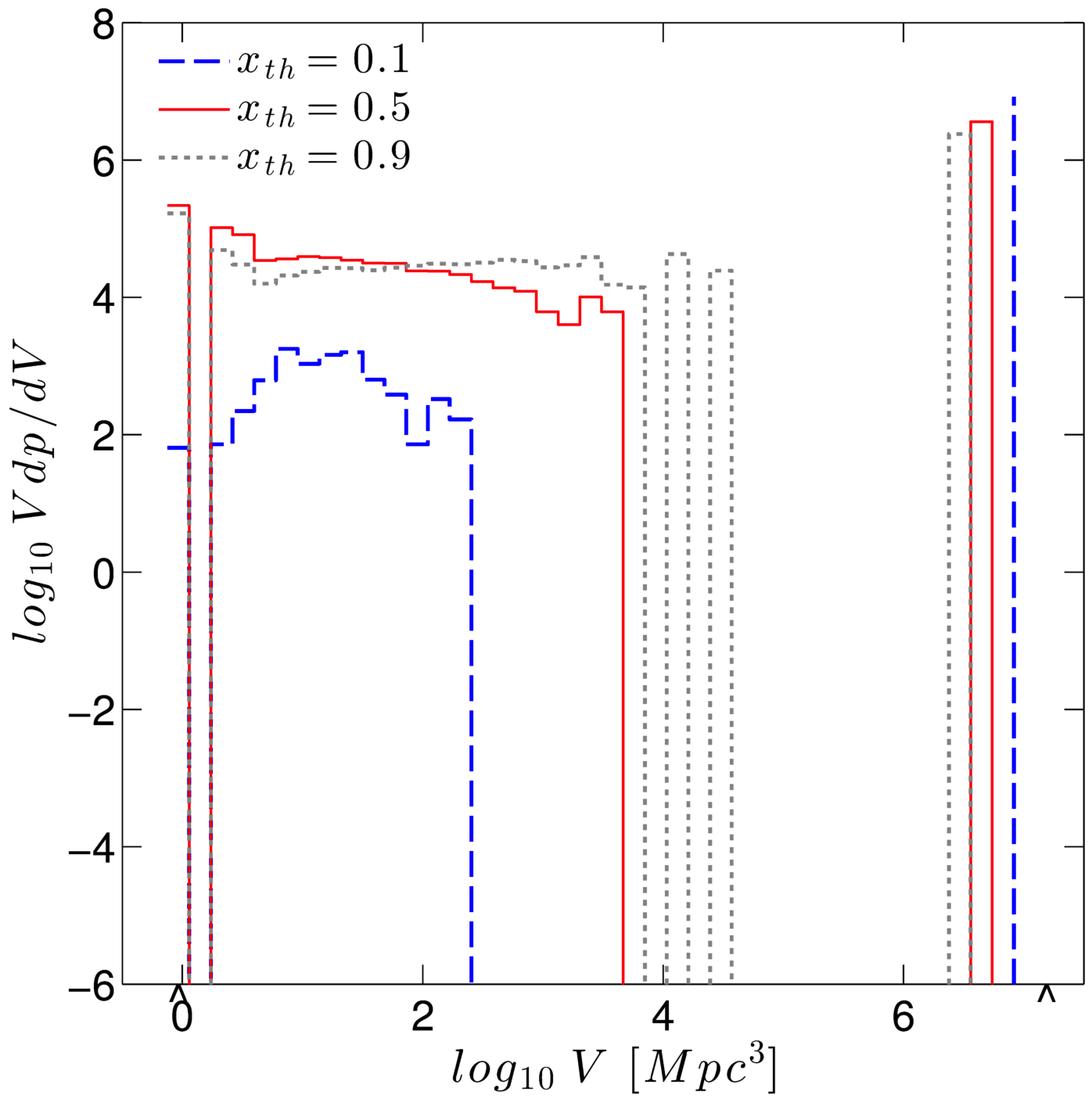}
\caption{The effects of three different thresholds, $x_{th} = 0.1, 0.5,$ and $0.9$, in the FoF method for the ionization simulation for the L1 (no LLS) case at z = 9.4 with global ionization fraction of $\langle x_m \rangle$ = 0.5. The two arrowheads on the abscissa, from left to right, mark the volume of a single cell ($0.25 \ \text{Mpc}^3$) and the volume of the box ($4.3 \times 10^6 \ \text{Mpc}^3$) respectively.}
\label{size_thresh_114}
\end{figure}

The three reionization histories are the result of available ionizing photons from the sources. Figure \ref{phot_atoms} shows the cumulative number of photons per baryons as a function of redshift. At the end of the ionization, the photons per atoms are twice as many for the LLS case. The LLSs therefore absorb approximately one extra photon before reionization is completed with not much difference between the two LLS cases. Their effect dominates over the recombinations included in the simulation which only consume 0.5 photon per baryon by the end of reionization.

\subsection {Morphology of H{\small{II}}  Regions}
The morphology of ionized regions is complex. We use several methods to qualitatively and quantitatively study the \hii region sizes, distribution, and evolution in our set of simulations. These methods, as discussed below, provide complementary information. 

One of the expected outcomes of different mean free paths $\ell(z)$ for the ionizing photons, is a changed evolution of the size distribution of the \hii regions and the consequent ionization fraction history. Once the \hii regions grow larger than the mfp in certain directions, not all sources inside the region can contribute to their growth and therefore they will not grow as fast as in the case without LLS. The \hii regions can still grow larger than the mfp because they are driven by many sources some of which lie closer than the mfp to the edge of the region.

%% Figure
\begin{figure*}
\centering 
\includegraphics[scale=0.52]{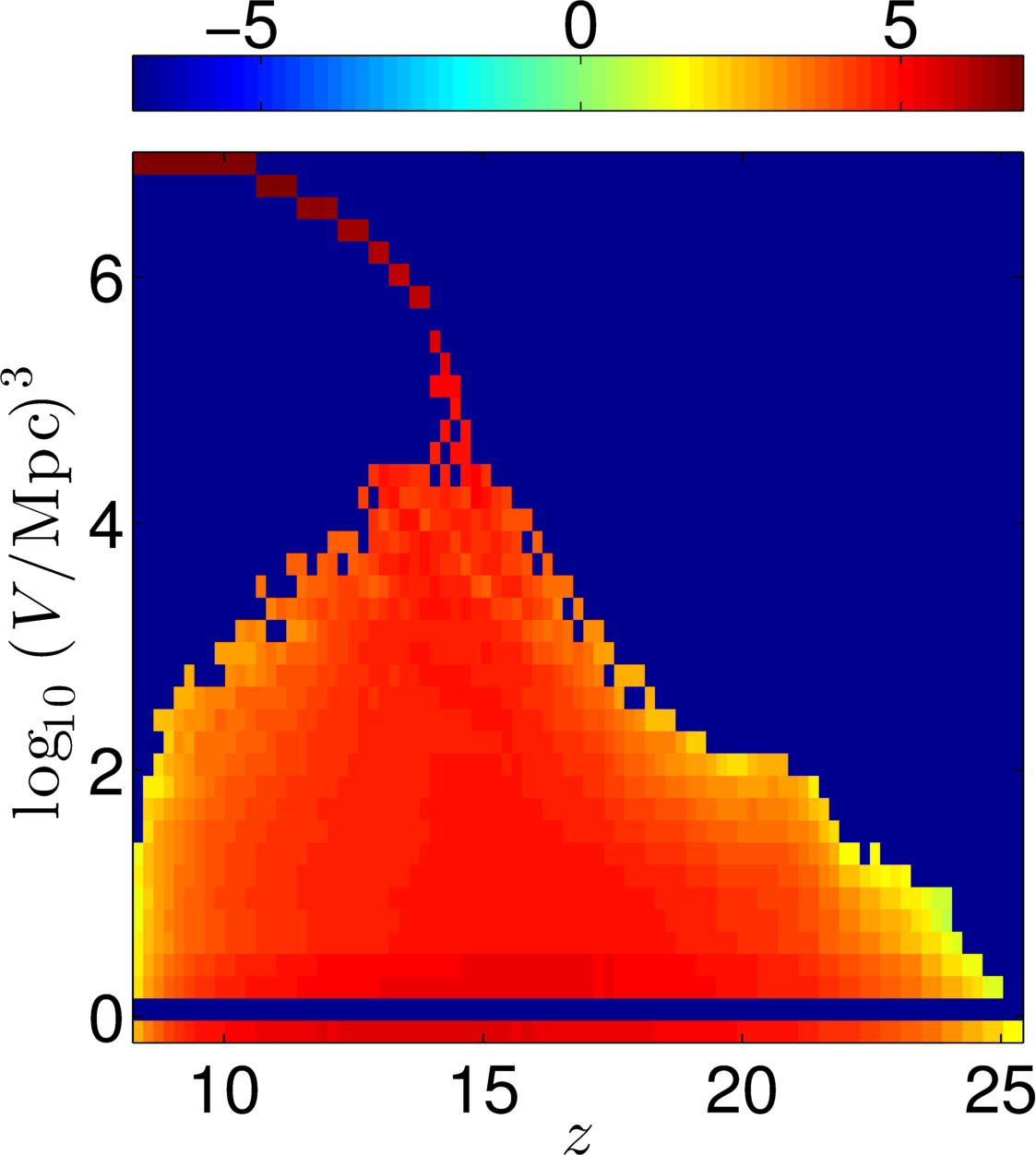}
\includegraphics[scale=0.52]{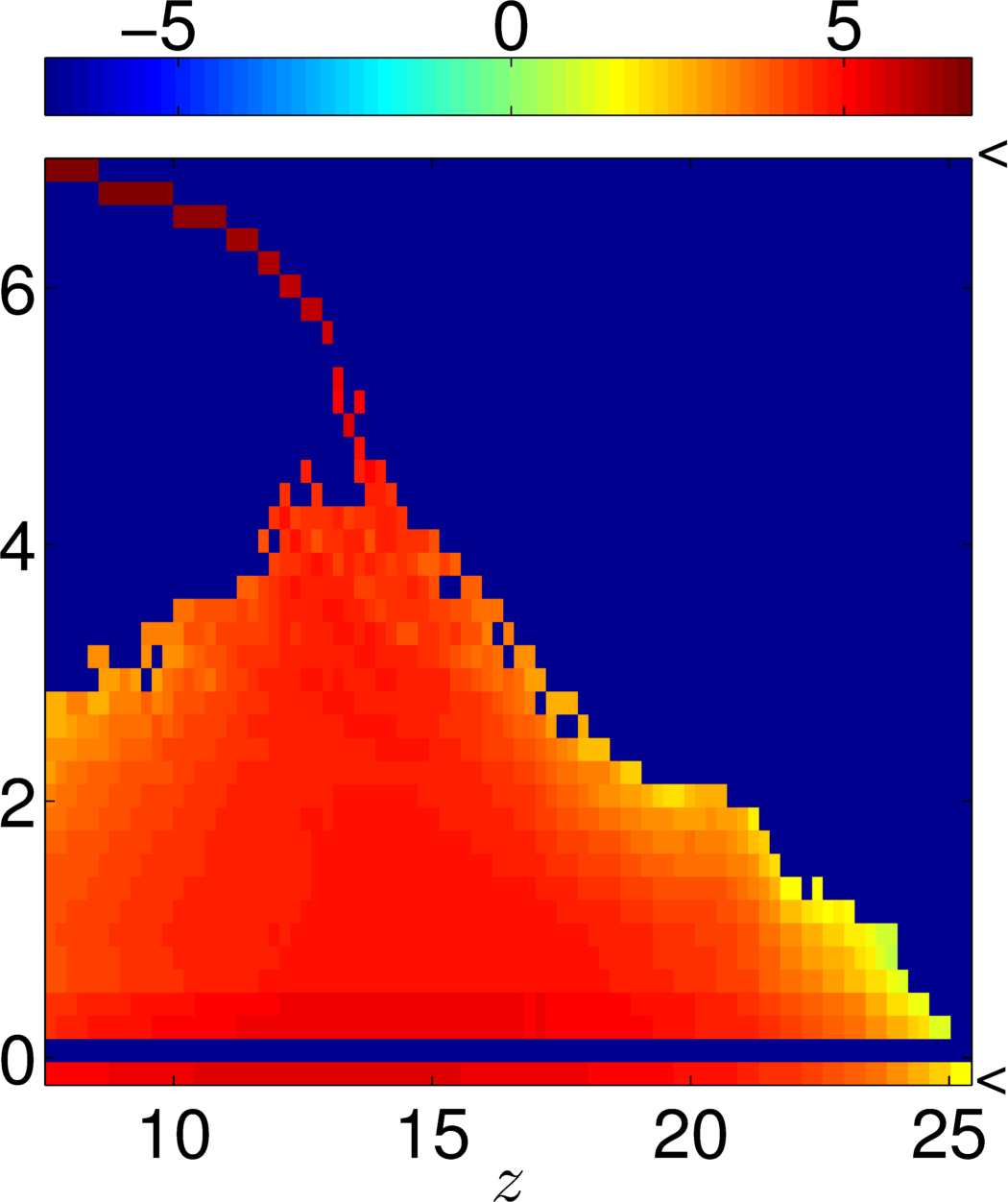}
\caption{Size distributions of the ionized H{\small{II}}  regions using the FoF method for the models L1 (left) and LLS2 (right) as a function of the redshift. The threshold used is $x_{th} = 0.5$. The colors correspond to the log$_{10}$ scale of the $V dp/dV$. The two arrowheads on the ordinate (right), from bottom to top, mark the volume of a single cell ($0.25 \ Mpc^3$) and the volume of the box ($4.3 \times 10^6 \ Mpc^3$) respectively.} 
\label{114_fof_im}
\end{figure*}

\subsubsection{Evolution of H{\small{II}} Regions}

Figure~\ref{xfrac-all} shows examples of the morphologies and growth of the ionized patches in the set of our simulations. The top panels show the ionization history of the fiducial model L1 (case without LLS) spanning redshifts, $z = 16.9 \mbox{-} 8.5$, approximately 370 million years. The panels from left to right show the slow ionization process, which reaches a mass weighted ionized fraction of 1\% only at  $z = 16.9$, even though the first halos in the simulation (with $M>10^8 M_\odot$) form at $z = 21$. 

The bottom three panels of Figure~\ref{xfrac-all} emphasize the morphological and topological difference between the three simulations. The three spatial ionization slices compare the global ionization of $\langle x_m \rangle= 50\%$ for models, L1, LLS1, and LLS2. The most obvious features are the different sizes of the larger \hii regions, especially between the non-LLS and the LLS models. This is indicative of the presence of LLSs slowing down the merger process. Between models LLS1 and LLS2 the differences in shapes and sizes of the ionized regions are not severe. However, in the detailed statistical analyses discussed below, some differences emerge. As expected, the slow growth of ionized regions delays the completion of reionization for the LLS simulations.

\subsection{Size Distribution of H{\small{II}} Regions } \label {sec-42}

In this section we quantify the results seen in Figure \ref{xfrac-all} by using three different methods to study the size evolution of the \hii regions in our numerical simulations. The statistical property measured in this analysis is the probability distribution function of the volumes (radii) of the \hii regions. The three approaches employed to estimate these size distributions are the friends-of-friends \citep[FoF, see][]{Iliev-2006aa}, the spherical average \citep[SPA, see][]{Zahn-2007} and 3D power spectra methods. All of these algorithms differ in their approach of defining the size of the \hii regions as discussed below. However, the different techniques complement each other and together provide a greater insight into the morphologies of the \hii regions
 \citep[for further details see][]{Friedrich:2011}.

\subsubsection {Friends-of-Friends}

The friends-of-friends (FoF) algorithm operates on the ionized fractions and generates a catalog of connected ionized \hii regions. For a chosen ionization threshold, $x_{th}$, the algorithm connects all the ionized neighboring cells and classifies them in a friendship based topology using the `equivalence class' or `sameness' method of the Numerical Recipes \citep{Press-1992}. The \hii regions catalogs based upon the ionization threshold and volume size are thus generated. These catalogs provide detailed insight in the evolution of the number densities and size distributions of the topologically-connected ionized regions. This method was first introduced in~\cite{Iliev-2006aa}.

Figure~\ref{size_thresh_114} shows the probability distribution function $V dp/dV$ at different friends-of-friends threshold values, $x_{th} = 0.1, 0.5,$ and $0.9$, versus the volume of the \hii regions for the L1 model at a global ionization fraction of $\langle x_m\rangle = 0.5$ ($z = 9.457$). The figure highlights the effects of the threshold. The lower threshold values reduce the number of smaller \hii regions and increase the number of the larger ones. The results of the FoF method therefore depend on the choice of the threshold \citep[as already shown in][]{Friedrich:2011}. However, for a given value of the threshold the results of different simulations can still be compared in a meaningful way.

The topological evolution of the \hii regions for the L1 and LLS2 models is shown in Figure~\ref{114_fof_im} for a threshold value of $x_{th} = 0.5$. The difference between the two LLS models was not discernible in the FoF analysis, therefore, only one LLS model is shown in Figure \ref{114_fof_im}. The colors represents the probability distribution $Vdp/dV$. It is  evident from Figure~\ref{114_fof_im} that the \hii regions grow as the ionization fraction increases up to a point where the distribution of volumes separates into two populations comprising of very large and relatively smaller sized regions. The emergence of the dichotomy is primarily due to the merging of smaller regions into larger volumes as the ionization fronts travel outwards from the higher density areas. As expected, the larger \hii regions of the order of $\sim 10^6$ Mpc$^3$ appear slightly earlier in the L1 ($z = 13.48$) case as compared to the LLS2 ($z = 12.31$) model. Another difference between the two models is that the largest of the ``small'' \hii regions disappear faster in the L1 case than in the LLS models. For the no LLS case L1, at $z = 8.34$ ($\langle x_m \rangle = 0.99$), the largest of the ``small'' \hii regions are 50\% smaller than the equivalent population in the LLS2 simulation. This is indicative of fewer mergers in the LLS cases as these regions only disappear when they merge with the larger regions. Towards the end of the ionization, the main contributors to the global average of the ionization fraction are the largest regions. This emphasizes the trends we have noticed earlier where the ionizing photons of the shorter mean free paths are absorbed and fail to contribute in the formation of \hii regions that grow and merge. 
From the observational perspective, the large \hii regions could be tomographically imaged with SKA-class interferometers. The volume distribution of such regions may help put limits on the mean free path and therefore on the LLS models.

%% Figure
\begin{figure}
\centering \includegraphics[scale=0.44]{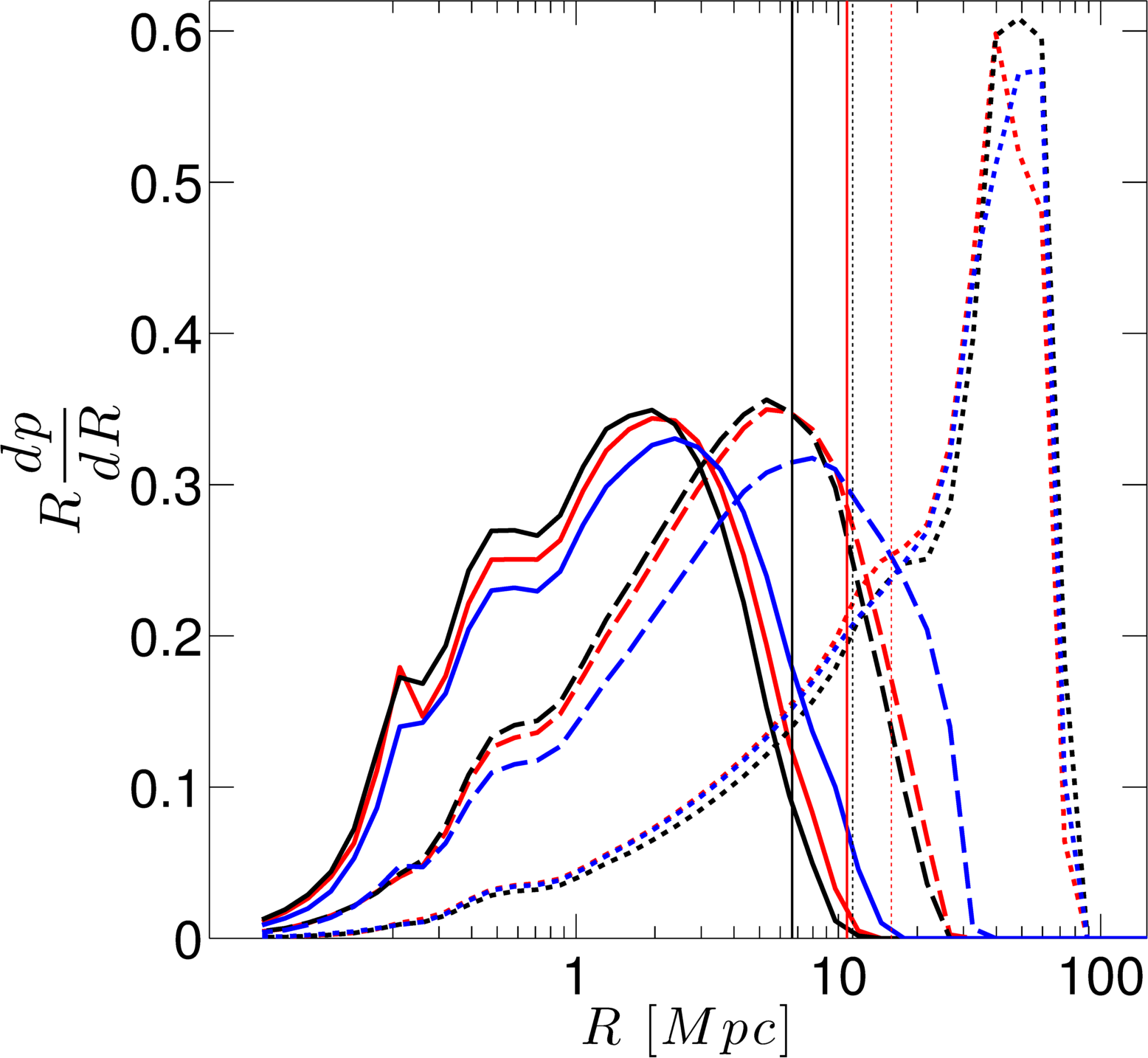}
\caption{Probability distribution function $R\ dp/dR$ per radial bins of spherical H{\small{II}}  regions as measured by the spherical averaging algorithm for the no LLS (L1, blue), and LLS models (LLS1, red; LLS2, black). The three sets are for a mass-weighted global ionization fraction $\langle x_m \rangle =$ 50\% (solid), 70\% (dashed), and 90\% (dotted). The threshold used is $x_{th} = 0.9$. The vertical lines correspond to the mean free path for the two LLS models as listed in Figure \ref{mfp} for 50\% and 90\% ionization fractions.}
\label{sphavg}
\end{figure}

\subsubsection {Spherical Averaging}

The second algorithm to evaluate the size distribution statistics was developed in~\cite{Zahn-2007}. This spherical averaging technique constructs spheres of varying radii around each 
cell in the ionization simulation and estimates the enclosed ionization fraction. The largest spheres with ionization fraction greater than the defined threshold, $x_{th}$, define the spherically averaged \hii regions. In contrast to the FoF method, the SPA technique yields a smoother and spherical distribution function, biased towards the shorter axes of triaxial structures.

The SPA analysis highlights a similar behavior in the evolution of the ionized regions as seen earlier with the FoF method. The shorter mean free path for the LLS simulation affects the growth as measured in the radii of the spherical regions. In Figure \ref{sphavg},  the solid, dashed, and dotted lines correspond to the 50\%, 70\%, and 90\% global ionization rates respectively. The color motif remains the same throughout the paper with L1 (blue), LLS1 (red), and LLS2 (black). 
The redshifts for which the distributions are calculated $z = 9.457, 8.958, 8.636$ for L1, $z =  9.236, 8.636, 8.172$ for LLS1, and $z = 9.164, 8.515, 8.012$ for LLS2. The vertical lines are the mean free path for the two LLS models as shown in Figure \ref{mfp}. These lines are plotted for the redshifts of 50 and 90\% ionization.

It is evident in Figure \ref{sphavg} that the number of regions with smaller radii ($< 0.2-0.3$ Mpc) is relatively similar for all models at different stages of the ionization history. However, for the larger radii ($< 0.3$ Mpc) differences in the probability distributions between L1 and both the LLS models emerge; especially as the ionization progresses. For example, at the 50\% and 70\% ionization stages, the radii for the peak probability in the L1 simulation are larger by a factor of $\sim 1.2$ compared to the LLS cases. To a lesser extent, the similar trend is seen when comparing the longer mfp LLS1 model to the LLS2. This is indicative of impeded ionization for the LLS models. At 90\% ionization most of the \hii regions have merged and therefore the distributions of spherical volumes are similar. In addition, at radii values reaching $\sim 80-90$ Mpc  the \hii regions are as big as the simulation box ($\text{R}_{114} = 81.42$ Mpc). The spherical averaging algorithm reaches the limits of its applicability at this stage. 

The maximum radii difference between L1 and LLS2 for 50, 70, and 90\% ionization varies from roughly 18, 33, to 18\%. Another characteristic that is apparent from the SPA analysis is that for early times (10\% ionization) smaller (radius $< 0.3$ Mpc) spherically averaged spheres contribute most to the probability distribution, $R\ dp/dR$. As the ionization progresses, the contribution from smaller spheres decreases and from the larger spheres increases indicating growing \hii regions and mergers. This trend is same for all the three models with impeded growth of radii for LLS models as evident in Figure \ref{sphavg}. The bubbles of larger radii merge earlier for the L1 model. When ionization reaches 50\%, the maximum radii of the \hii regions are comparable to the mfp, more so for the LLS2 than for the LLS1 model. However, as the ionization progresses, the radii of the \hii regions grow beyond the mfp due to mergers.

%% Figure
\begin{figure}
\centering 
\includegraphics[scale=0.4]{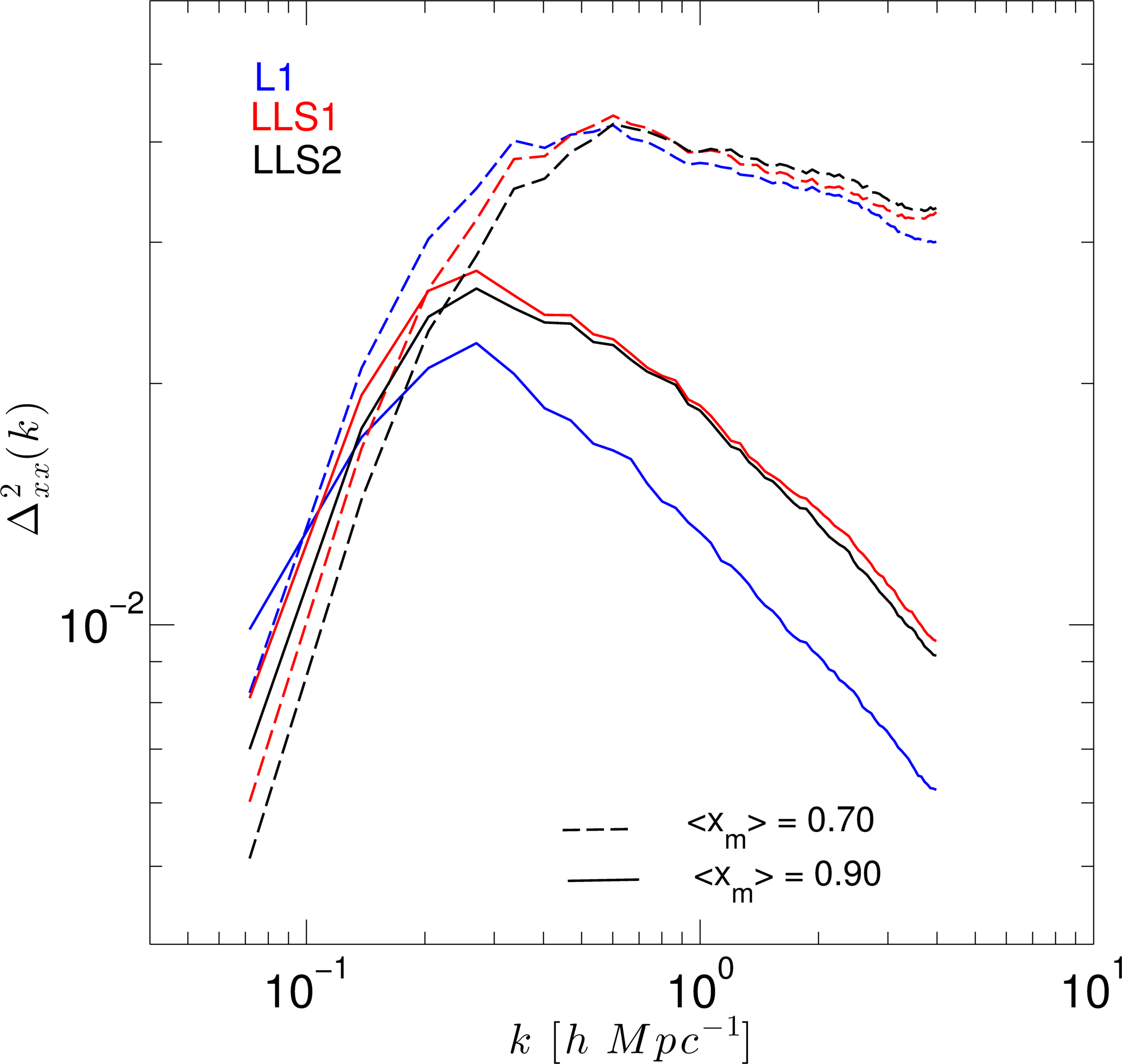}
\caption{The log-log plot of the dimensionless 3D power spectra of the ionized fraction, at $\langle x_m \rangle = 70\% $ and $\langle x_m \rangle= 90\% $ for the three models.}
\label{fig:psxfrac}
\end{figure}

%% Figure
\begin{figure*}
\centering
 \includegraphics[width=3.1in]{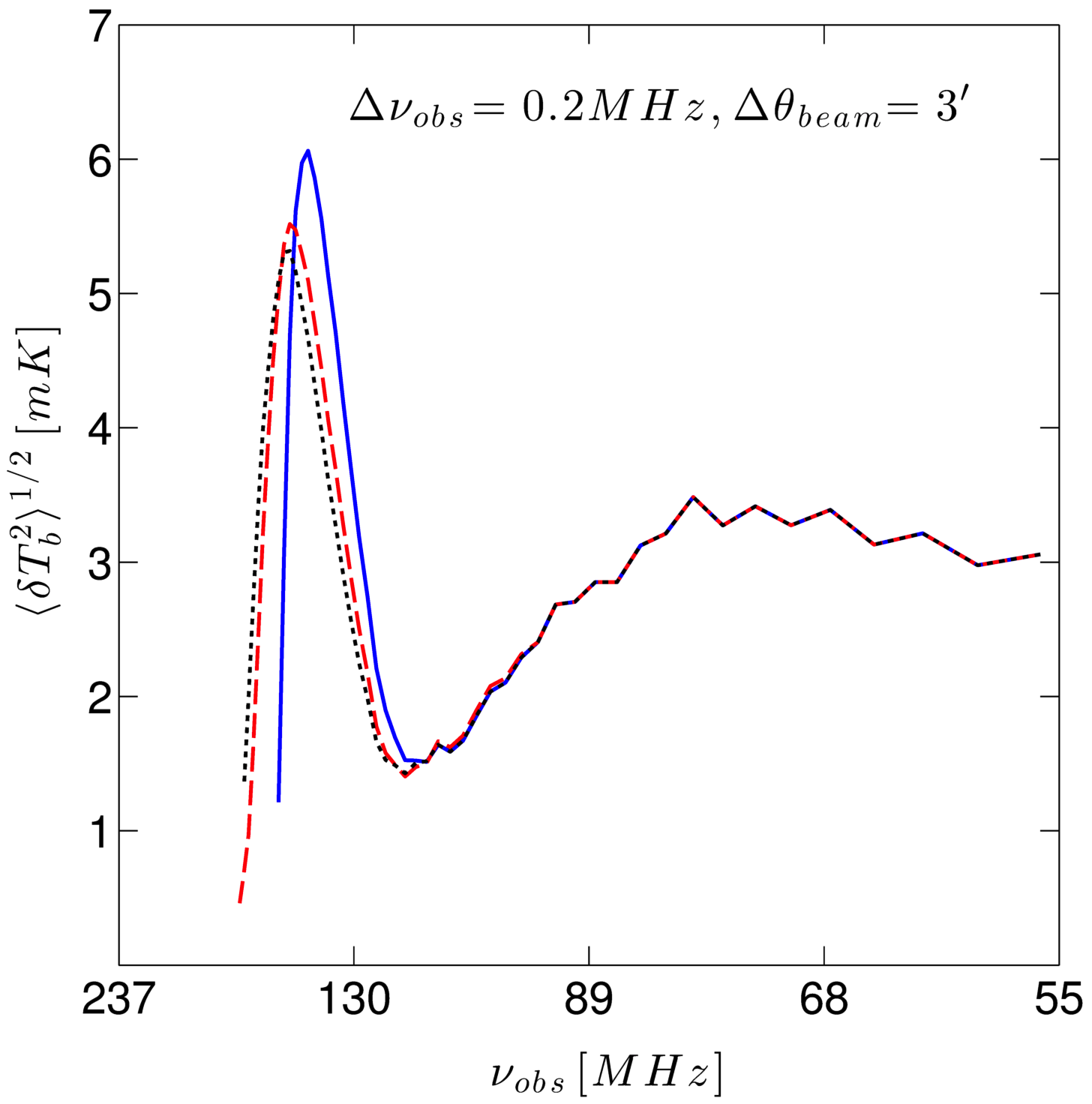}
 \includegraphics[width=3.1in]{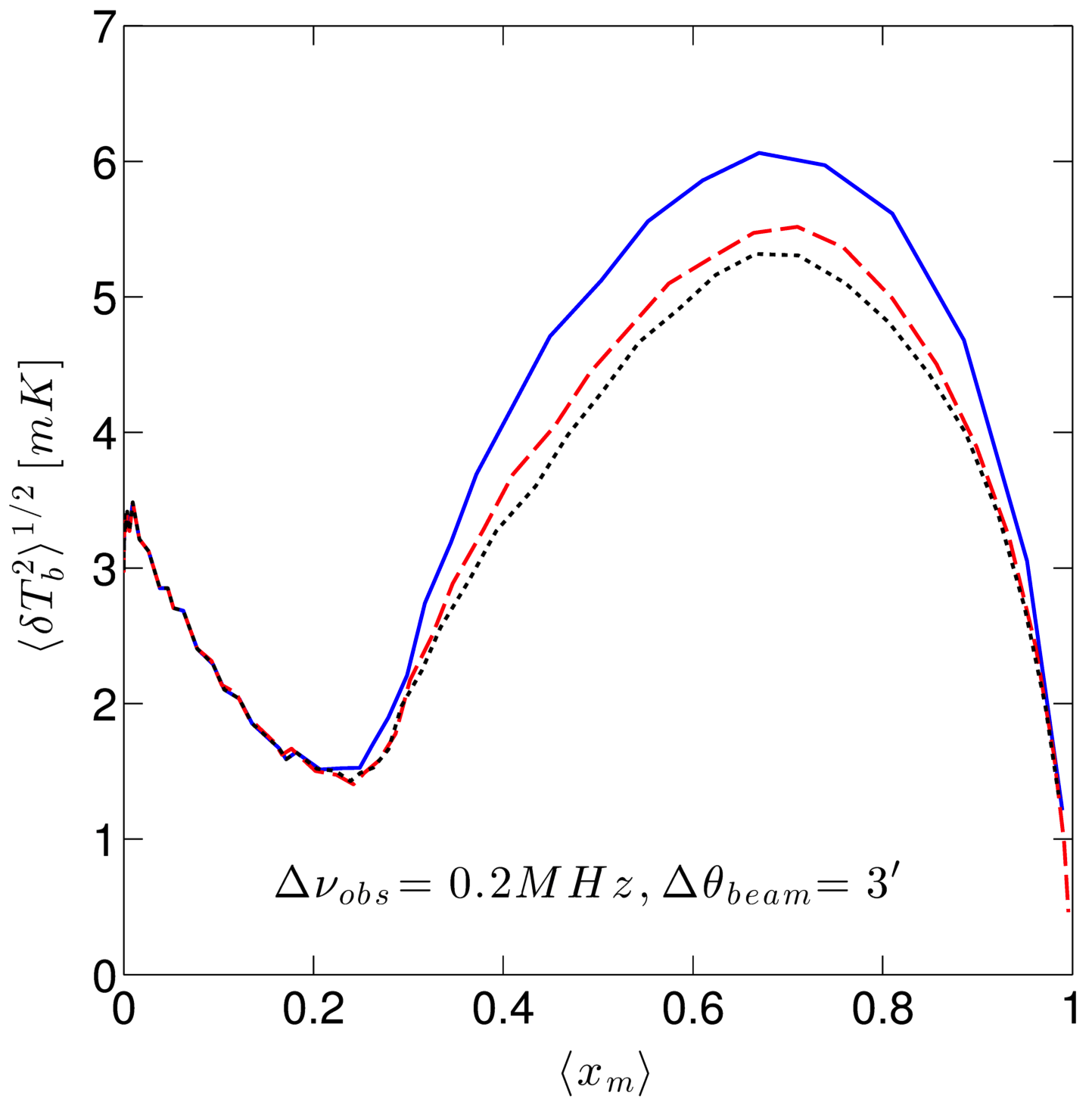}
\caption{ The evolution of the RMS fluctuations of the 21-cm background, for beamsize 3\arcmin \ and bandwidth 0.2 MHz and boxcar filter vs. frequency (left) and vs. average ionization (right). The simulations shown are L1 (blue, solid), LLS1 (red, dashed), and LLS2 (black, dotted).}
\label{fig:rms}
\end{figure*}

These results are consistent with ones from FoF method. The unimpeded ionizing photons in the L1 model noticeably differentiate the ionization and bubble growth history from that of the LLS models. There is not much difference between the LLS ionizing histories themselves, with radii less than 10\% different at different stages of the ionization rendering them hard to distinguish.

%% Figure
\begin{figure*}
\centering \includegraphics[scale=0.3]{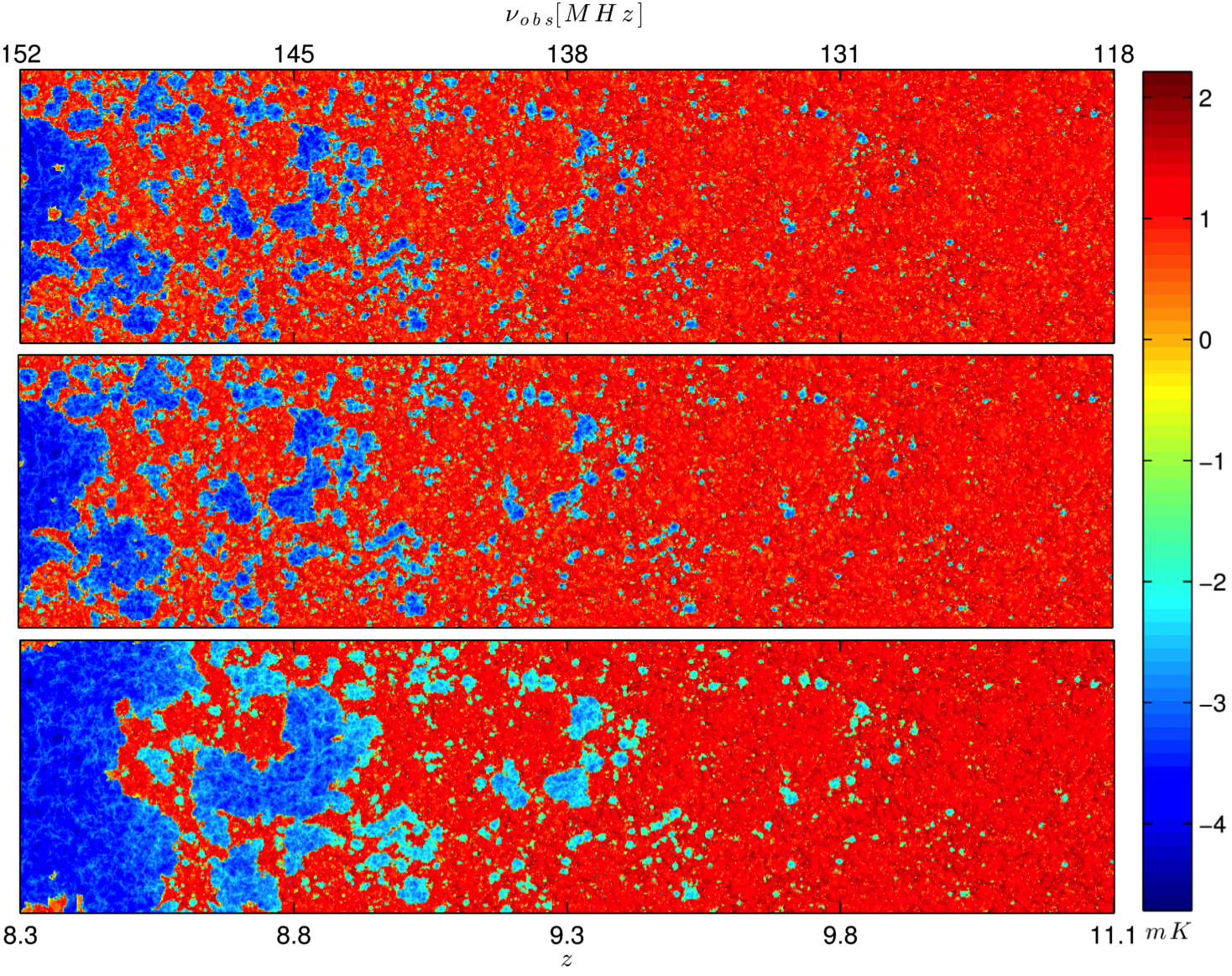}
\caption{From the top are the three images from simulations LLS2, LLS1, and L1 for the box size 114 \mpch. The images depict the position-redshift/frequency ionization brightness temperatures. The abscissae of the images correspond to redshift range from $z = 8.3 \mbox{-} 11.1$ and equivalent observational frequencies ($\nu_{obs} \ [MHz]$). The ordinates of the images are the comoving spatial dimension of 114 \mpch \ and 4 times that in the abscissa. The images show the differential brightness temperature in log scale ($log_{10} \ \delta T \ [mK]$) at the full grid resolution. The images are corrected for redshift-space distortions due to the peculiar velocities.}
\label{slices}
\end{figure*}

\subsubsection {Ionized fraction Power Spectrum}

The third method for the volumetric analysis is the power spectrum of the ionized fraction field. The power spectrum measures the contribution from different spatial frequencies and therefore is sensitive to the underlying structures. We calculated the dimensionless power spectrum per comoving wavenumber $\Delta^2_{xx} (k) = k^3 P_{xx}(k) /2\pi$, where $k$ [\hmpc] is the wavenumber and $P_{xx}(k)$ is the spherically averaged square of the absolute value of the 3D Fourier transform of the ionized fraction in coeval volume. 

Figure \ref{fig:psxfrac} shows the power spectra for tow ionization stages, namely, at 70\% and 90\%. At the 70\% ionization stage, the largest difference between the L1 and LLS cases is for wavenumbers smaller than $\sim 0.3$ \hmpc \ where L1 has approximately 1.6 times more power than LLS2 and 1.3 times more for LLS1. This is indicative of the presence of more structures on larger scales for L1. For smaller scales the differences are small. As reionization reaches 90\%, L1 has $\sim 1.5$ times less power at wavenumbers larger than $\sim 0.3$ \hmpc \ and 1.4 time more power around $k\sim 0.1$ \hmpc, implying that the LLS models retain more small scale structure and have somewhat less large scale structure in the ionization field.

%-----------------------------------------
% Section 5 - OBSERVATIONAL SIGNATURES
%-----------------------------------------

\section{Observing Redshifted 21-cm} \label {sec-5}

Measurements of bubble sizes and shapes will require high signal to noise images at scales of a few arcminutes. This will be possible with the future Square Kilometre Array (SKA). The first generation of radio telescopes will instead focus on measuring statistical quantities of the 21-cm signal, such as the power spectrum. The discussion below explores the 21-cm signal using the variance of the brightness temperature fluctuations and 21-cm power spectra. In addition, we also show the morphology of the
21cm signal in the so-called light cone slices. 

\subsection {Variance of the 21-cm background}

The differential brightness temperature of the redshifted 21-cm emission with respect to the CMB is given by the spin temperature, $T_\text{S}$, of the neutral hydrogen and its density, $\rho_\text{H\tiny I}$, and in the limit of $T_\text{S} \gg T_\text{CMB}$, is given by \citep{Field-1959},

\beq
\delta T_b = \frac{T_\text{S} - T_\text{CMB}}{1+z} (1-e^{-\tau})
\eeq

where, $z$ is the redshift, $T_\text{CMB}$ is the temperature of the CMB radiation at $z$, and $\tau$ is the corresponding 21-cm optical depth.

As seen in the previous analyses, the overall effect of the LLS is to slow down the ionization process and impede the growth of the \hii regions. This should manifest itself as two observable properties in the variance of the fluctuations. Firstly, the peak of the brightness temperature fluctuations for the LLS simulations should be delayed and therefore should be visible at relatively higher frequencies. Secondly, the amplitudes of the peaks should be diminished due to the relatively smaller size of the \hii regions in the LLS simulations. 

Figure~\ref{fig:rms} shows the evolution of the RMS fluctuations of the mean differential brightness temperature for the three simulations as convolved with LOFAR-like boxcar beam of size $3^\prime$ at a frequency bandwidth of 0.2 MHz. As depicted in the figure, at lower frequencies (early times) the fluctuations for all the three models are similar. The temperature fluctuations peak at 141 MHz for the L1 model and 147 MHz for the LLS1 and LLS2 models. The peak value of the brightness temperature RMS for the L1 model is 6.06 mK with the brightness temperature of 16.66 mK. The RMS is lower by about 9\% and 8.7\% for LLS1 and LLS2 models respectively. The RMS of the temperature fluctuations vs. the mass-weighted ionization global average is shown on the righthand side of Figure~\ref{fig:rms}. The brightness temperature fluctuations are again seen following each other very closely at early times. However, as ionization reaches 20\% the temperature fluctuations for different models start to diverge and peak at about 65-70\% ionization. After this the temperature fluctuations decrease and are indistinguishable as reionization completes.

As mentioned earlier, these differences in brightness temperatures are manifested by the varying distribution and growth of \hii regions in the different models as seen in the statistical analyses of previous sections. These fluctuations have been averaged over by a LOFAR-like beam and bandwidth. This is a simple first order estimate. Detailed and more accurate estimates require defining a noise budget including system temperatures, gains and phase errors, along with propagation effects (foregrounds, ionosphere etc.) and telescope based visibility sampling functions, see \emph{e.g.} \citet{Patil-2014}. Expectedly, increasing bandwidth reduces the RMS as the fluctuations are smoothed out for a wider bandwidth. Increasing of the resolution of the beam increases and broadens the RMS. This is also expected as a smaller beam is sensitive to small scale fluctuations that are smoothed out by larger beams. Similar to the analyses in the previous sections, the differences are more pronounced between non-LLS and LLS models. However, based solely upon brightness temperature fluctuations it will be non-trivial to distinguish between the LLS models, see Figure \ref{mfp}.

\subsection {Evolution of the patchiness}

Figure~\ref{slices} shows slices through the simulation cubes along the redshift (frequency) axis, also known as light cone slices. The ionized fraction of the simulation cubes is converted to the 21-cm emission differential brightness temperature, shown in log scale in mK, for the three models, shown from the top, L1, LLS1, and LLS2. For the desired range of redshifts, the data from the cubes is interpolated along the redshift/frequency axis. The interpolation is performed along the plane with an oblique angle of $10^\circ$ across the cubes in order to observe different structures along a random line of sight. The neutral regions are shown in red and the \hii regions cover the dynamic range through blue as shown by the color bar of Figure~\ref{slices}. No corrections for redshift space distortions are applied.

From Figure~\ref{slices} it is evident that at high redshifts the \hii regions are small and distributed sparsely. These regions closely trace the ionizing halos. The effect of the different mfps of the the three models on the ionization becomes visually evident at redshift $z \sim 9.8$ increasing with lower redshifts. The \hii regions for the three models grow and merge at different paces. For this reason by redshift $z \sim 8.3$ the L1 model is fully ionized, whereas, for the models with LLS, the mass weighted global average ionization for LLS1  is $\langle x_m \rangle  \sim 83\%$ and LLS2, $\langle x_m \rangle  \sim 78\%$. The spatial axis of the coeval simulation box at redshift $z = 9.457$ subtends an angle of $\sim 0.97^{\circ}$ in the sky with each pixel of size $13.76''$. Without any astrophysical and instrumentational propagation effects the \hii regions at lower redshifts are large enough to be directly observed by arrays with $\sim 1'$ angular resolution capabilities. The effects due to the synthesized beam smooth the fine reionization structure but the statistical measurement of the signal is still achievable. 

%% Figure
\begin{figure}
\centering \includegraphics[scale=0.44]{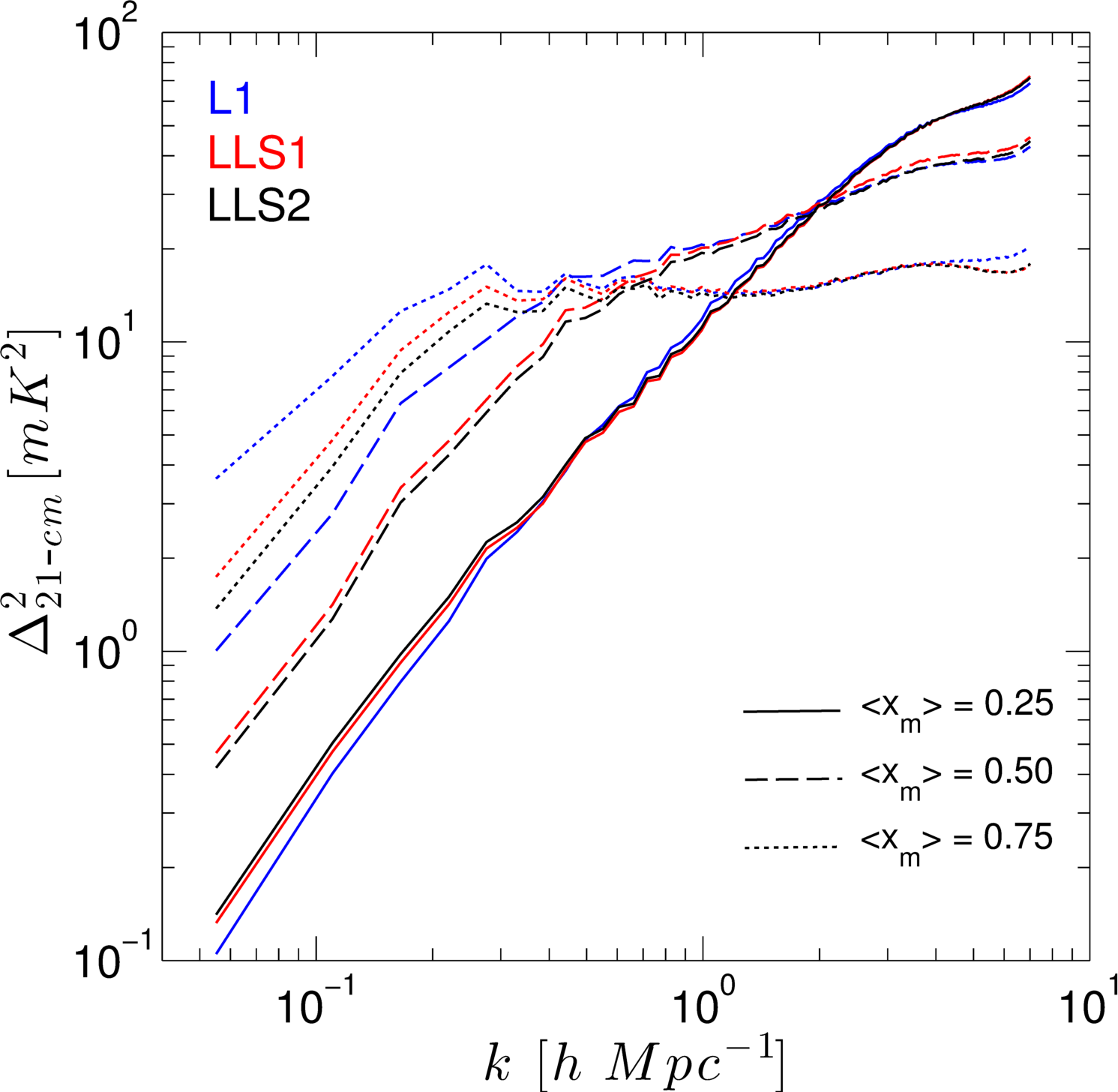}
\caption{The 3-D spherically averaged 21-cm differential brightness temperature fluctuation power spectra for models L1 (blue), LLS1 (red), and LLS2 (black) at mass weighted global ionization of 25\% (solid lines), 50\% (dashed lines), and 75\% (dotted lines).}
\label{ps-1}
\end{figure}

\subsection {Power Spectrum of 21-cm}

The power spectrum of the differential brightness temperature distribution, $\delta T_b$, is defined as,

\begin {equation}
\langle \widetilde{\delta T_b^\star}({\bf k}) \widetilde{\delta T_b}({\bf k'}) \rangle
	= (2\pi^3) P_{21} ({\bf k}) \delta_D^{(3)}({\bf k - k'})
\end {equation}

where, $\widetilde {\delta T_b}$ is the Fourier transform of the differential brightness temperature, $\widetilde {\delta T_b^\star}$ is the complex conjugate, $P_{21}$ is the spherically averaged power spectrum, and $\delta_D^{(3)}$ is the three-dimensional Dirac delta function representing the sampling function of the Fourier transformed quantity. The power spectrum is in units of mK$^2$ and is also used in dimensionless form as,

\begin{equation} 
\Delta^2_{21\mbox{-}cm} ({\bf k}) = \frac {k^3}{2\pi^2} P_{21} ({\bf k}) \quad [\rm mK^2]
\end{equation} 

%% Figure
\begin{figure}
\centering \includegraphics[scale=0.44]{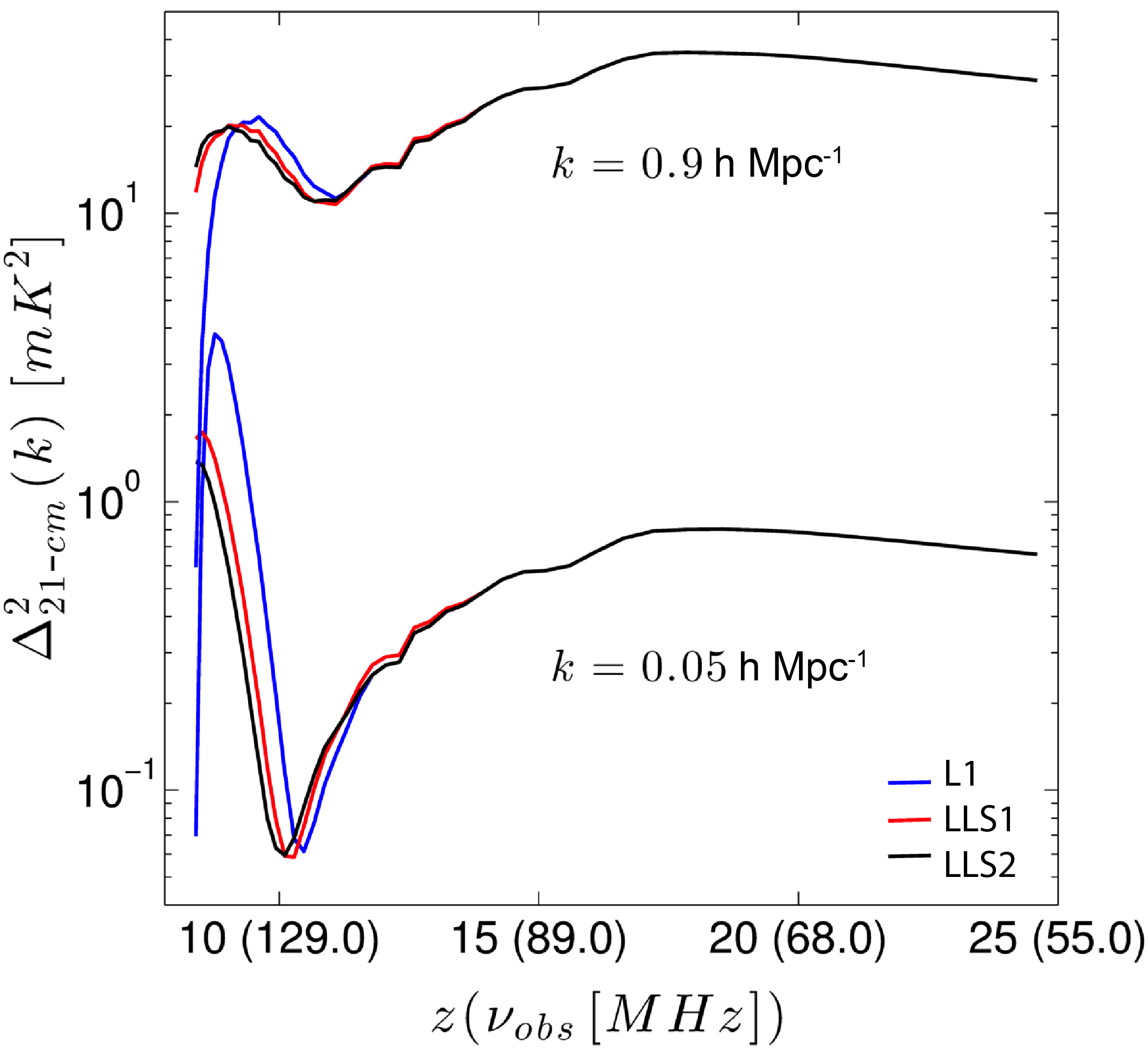}
\caption{The time evolution of the two $k$-modes, $k$ = 0.05 and 0.9 \hmpc \ of the spherically averaged 3-D power spectra of the differential brightness temperature fluctuations for the three models.}
\label{ps-kevol}
\end{figure}

%%Figure
\begin{figure*}
\centering \includegraphics[scale=0.6]{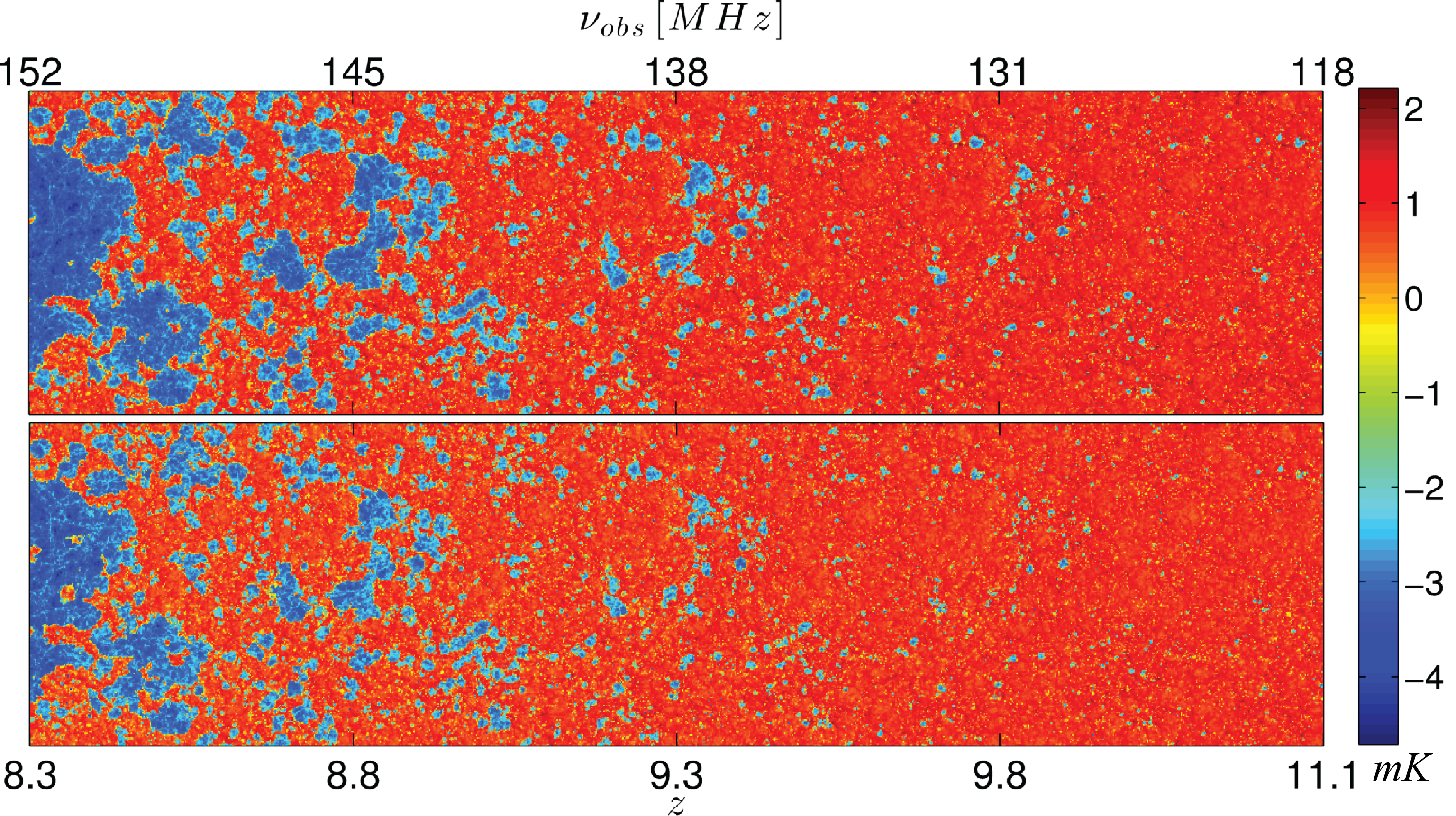}
\caption{Top - Model LLS1, Bottom - Model LLS1-PD. Both for $z = 8.34 - 11.08$.}
\label{posdep}
\end{figure*}

There are various schemes for estimating the power spectrum. In this paper, results from the \emph {mesh-to-mesh real-to-redshift-space mapping} (MM-RRM) methodology \citep{Mao-2012} are used. The MM-RRM uses the ionization fraction, density, and velocity data as input to estimate the power spectrum. This methodology takes into account the effects of redshift-space distortions and predicts accurate estimates of the 21-cm background with the caveat that at $k_N^{(256)} / 4 > 1.75 ^{-1}$ \hmpc \ the errors in the estimated PS are large.

The measurement of the power spectrum lends itself naturally to the radio interferometric observations since the visibilities of the interferometric measurement are sampling the Fourier transform of the sky and the power spectrum is the Fourier transform of the two-point correlation function. 

Figure \ref{ps-1} shows the dimensionless 21-cm  differential brightness fluctuation power spectra of the three models at three representative stages, namely, at global ionization average of 25\%, 50\%, and 75\%. The most distinct characteristic visible in the figure is that the fluctuations in the brightness temperature at large scales ( $k < 0.2$ \hmpc) grow by roughly 1.5 orders of magnitude for the fiducial L1 case and less than an order of magnitude of the LLS cases. This is a signature of the larger \hii regions causing larger temperature fluctuations. The smaller fluctuations, on the other hand, flatten out as reionization progresses. Another noteworthy feature is the divergence of the fiducial L1 model from the LLS models at the largest scales traced by the simulations. 

Figure \ref{ps-kevol} shows the evolution of the power spectra with redshift for the three models for two $k$ values, 0.05 and 0.9 \hmpc \ representing large and small scale fluctuations respectively.

The features in the 21-cm power spectra are consistent with the prior analyses. The recurring theme in the analysis of size and distribution of the \hii regions is that for the LLS models the growth of ionized regions is obstructed and ionization is delayed. This is well captured in the 21-cm power spectra, defining implications for the upcoming experiments. For both the scales ($k = $ 0.05 and 0.9 \hmpc) the observational frequency range from 140-150 MHz is where signals peak and the models are most differentiable. For large-scales ($k = 0.05$ \hmpc) the signal also goes through a minimum in the 123-127 MHz range with the lowest signal occurring for LLS2 followed by LLS1 and L1 models. While at this minimum the LLS models are only 15\% lower than L1, a more significant difference occurs at the maximum where LLS1 and LLS2 show a 27\% and 31\% decrease compared to L1. As can be noted, the difference between the two models is 0.63 mK$^2$. For small-scales ($k = 0.9$ \hmpc) the spectra peak very early on at $z = 17.85$ corresponding to a frequency of 75 MHz. This is region of the power spectrum where the models are virtually identical, but where they are also more likely to be affected by fluctuations in the spin temperature. The next peak for the small-scale features tracks the large-scale but occurs slightly earlier around 133 MHz with the corresponding dip occurring at 117 MHz. 

It is discernible from the power spectra that the contribution at small- and large-scales lag in the ionization process due to the reduction in the mfp caused by the dense LLSs. The delay in the overall ionization is also evident in the figure with the peak for the non LLS case rising at earlier times followed by models LLS1 and LLS2.

%-----------------------------------------
% Section 6 - Position Dependent LLS
%-----------------------------------------

\section{Spatially-varying LLS optical depth} \label {sec-6}

In this section we examine the case of spatially-varying, or position dependent LLS distribution for the Songaila model (LLS1), titled LLS1-PD, and compare it with the corresponding homogeneous distribution case. The code implementation for both cases was discussed in \S \ref{sect:algol}. 

By construction, the average LLS optical depth and the number of ionizing photons emitted at any one time is the same in the two cases. This yields essentially identical evolutions of the number of photons per atom and very similar mean reionization histories (not shown). However, the \hii region morphology is affected by the differing LLS distributions. In the spatially-varying case the LLSs are concentrated in the vicinity of the ionizing sources, acting like a type of a screen around them. This slows the growth of the ionized patches in the LLS1-PD case, resulting in more numerous, but smaller \hii regions early on than are found in the homogeneous case (Fig.~\ref{posdep}). In the former case there are also more partially or fully neutral gas pockets inside the ionized regions. At late times the larger ionized patches are more fragmented, with more structure. 

%%Figure
\begin{figure}
\centering \includegraphics[scale=0.42]{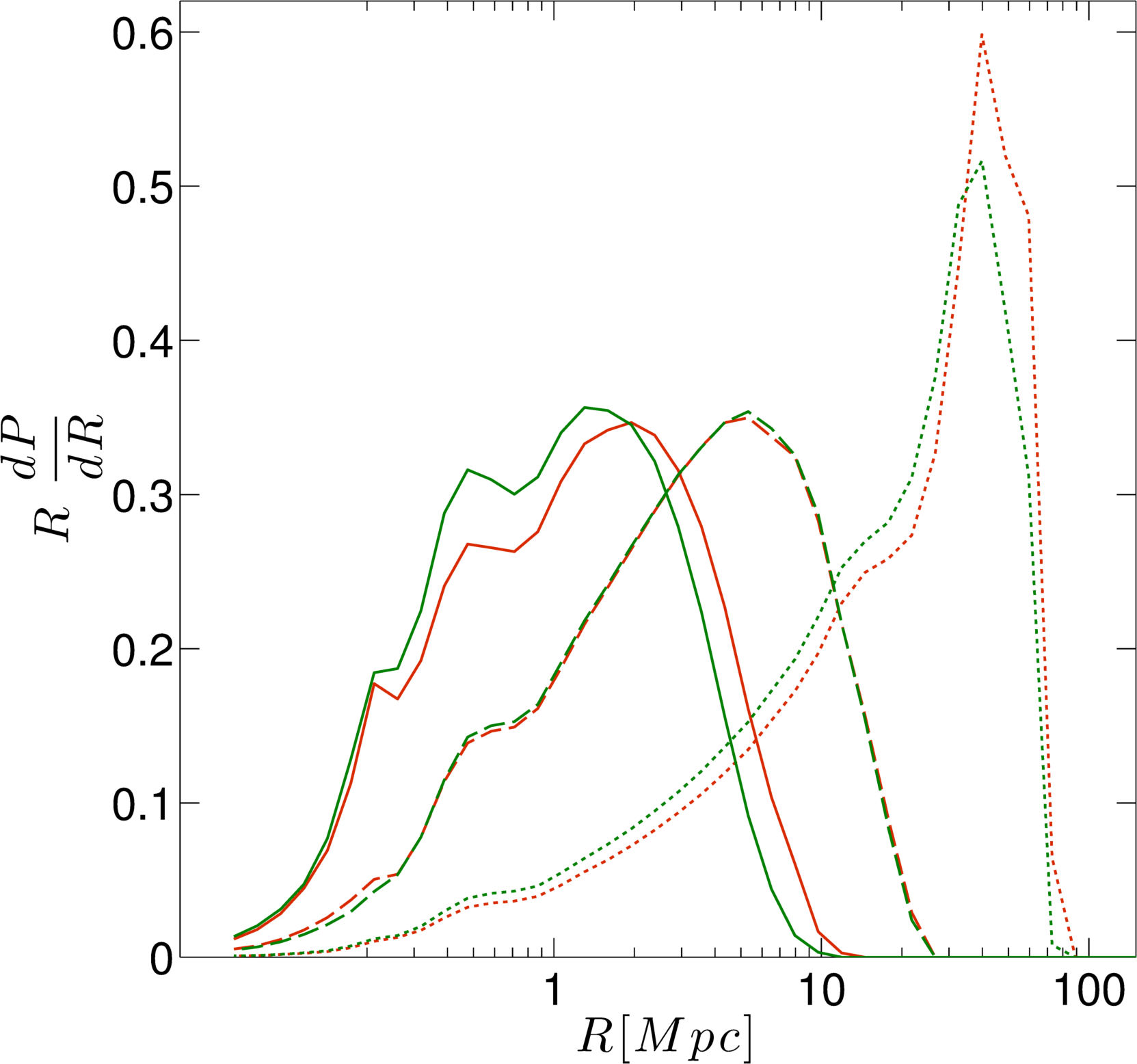}
\caption{Probability distribution function $R\ dp/dR$ per radial bins of spherical HII regions as measured by the spherical averaging algorithm for the LLS models (LLS1, red) and (LLS1-PD, green) The three sets are for the mass-weighted global ionization fraction $\langle x_m \rangle$ = 50\% (solid), 70\% (dashed), and 90\% (dotted). The threshold used is $x_{th} = 0.9$.}
\label{fig:spa}
\end{figure}

%% Figure
\begin{figure}
\centering \includegraphics[scale=0.4]{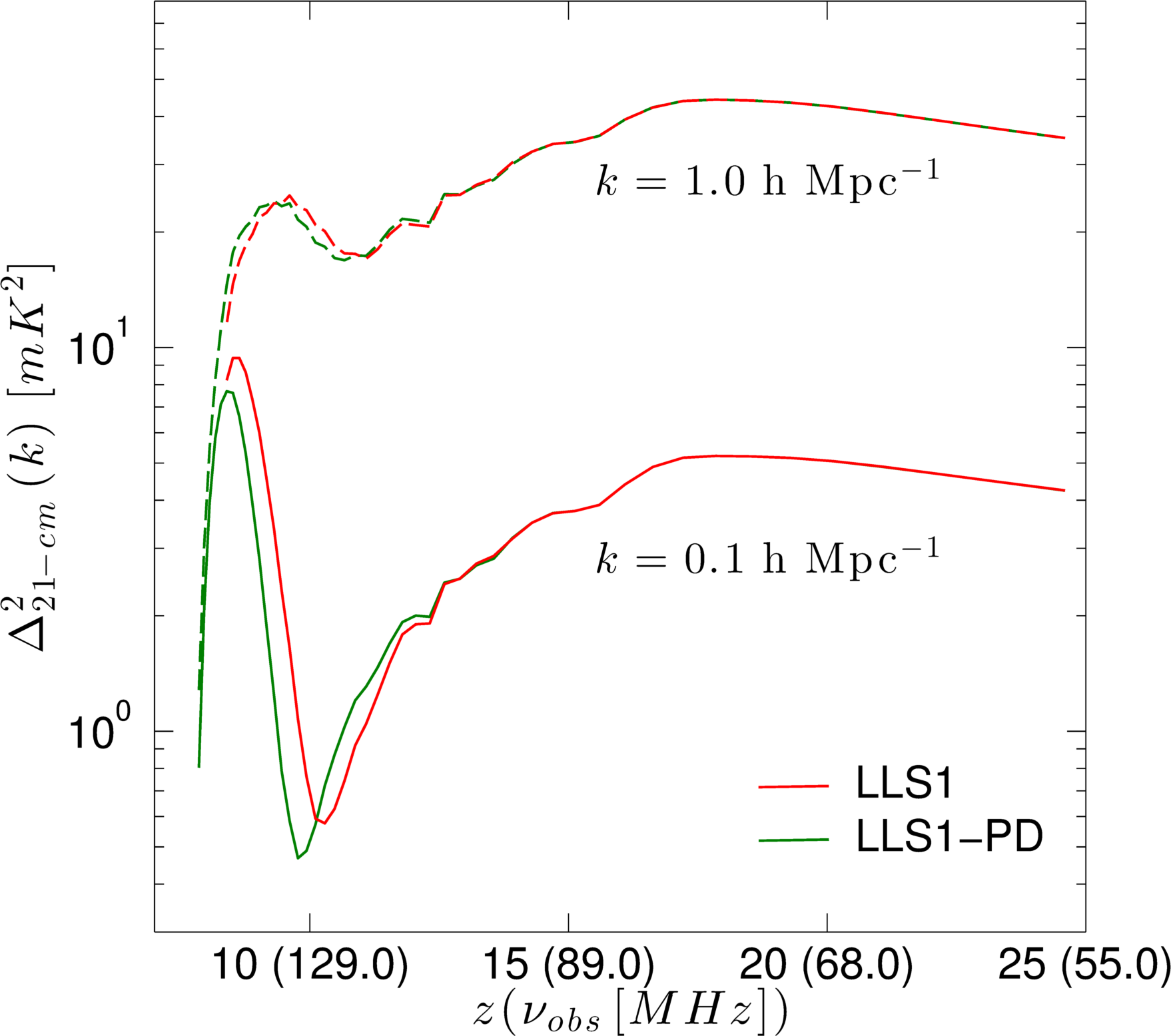}
\caption{The time evolution of the two $k$-modes, $k$ = 1.0 and 0.1 h Mpc$^{-1}$ of the spherically averaged 3-D power spectra of the differential brightness temperature fluctuations for the LLS1 (red) and LLS1-PD (green) models.}
\label{fig:psevol}
\end{figure}

These trends in the \hii region sizes can be quantified using the the spherical averaging (SPA) technique (Figure~\ref{fig:spa}). At $\langle x_m \rangle = 50\%$ for LLS1-PD the whole distribution is shifted noticeably to smaller sizes, with the peak probability distribution of the radii of the \hii regions is less than by a factor of 2. In addition, what becomes more evident in the SPA analyses is the growth pattern. At $\langle x_m \rangle = 70\%$, the probability distributions of LLS1 and LLS1-PD are virtually indistinguishable. This is an artifact of how the spheres are fit to the \hii regions in the SPA technique, which among other things erases the small-scale structure. However, as ionization progresses to $\langle x_m \rangle = 90\%$ the LLS1-PD model again shows on average smaller regions, although still a very similar PDF shape. The lags of the peak amplitudes of PDF diminishes for both the models as the ionization progress. This indicates that the most typical size of the \hii regions is almost the same, but there are still more small ionized patches in the LLS1-PD case, while the largest ones are smaller than in LLS1. 

These changes in the \hii region sizes should also be reflected in the 21-cm brightness temperature fluctuations. In Figure \ref{fig:psevol}, we show the evolution of two $k$-modes of the 21-cm power spectra, $k=1$ \hmpc \ and  $k=0.1$\hmpc, corresponding to small and large scale fluctuations, respectively. For both modes the peak fluctuations for LLS1-PD are shifted to slightly later times compared to LLS1. The differences are more pronounced for large \hii regions ($k = 0.1$\hmpc), with the peak lower by $\sim 1.8$ mK$^2$ and shifted by $\Delta \nu = 3.7~$MHz.

%% Figure
\begin{figure}
\centering \includegraphics[scale=0.44]{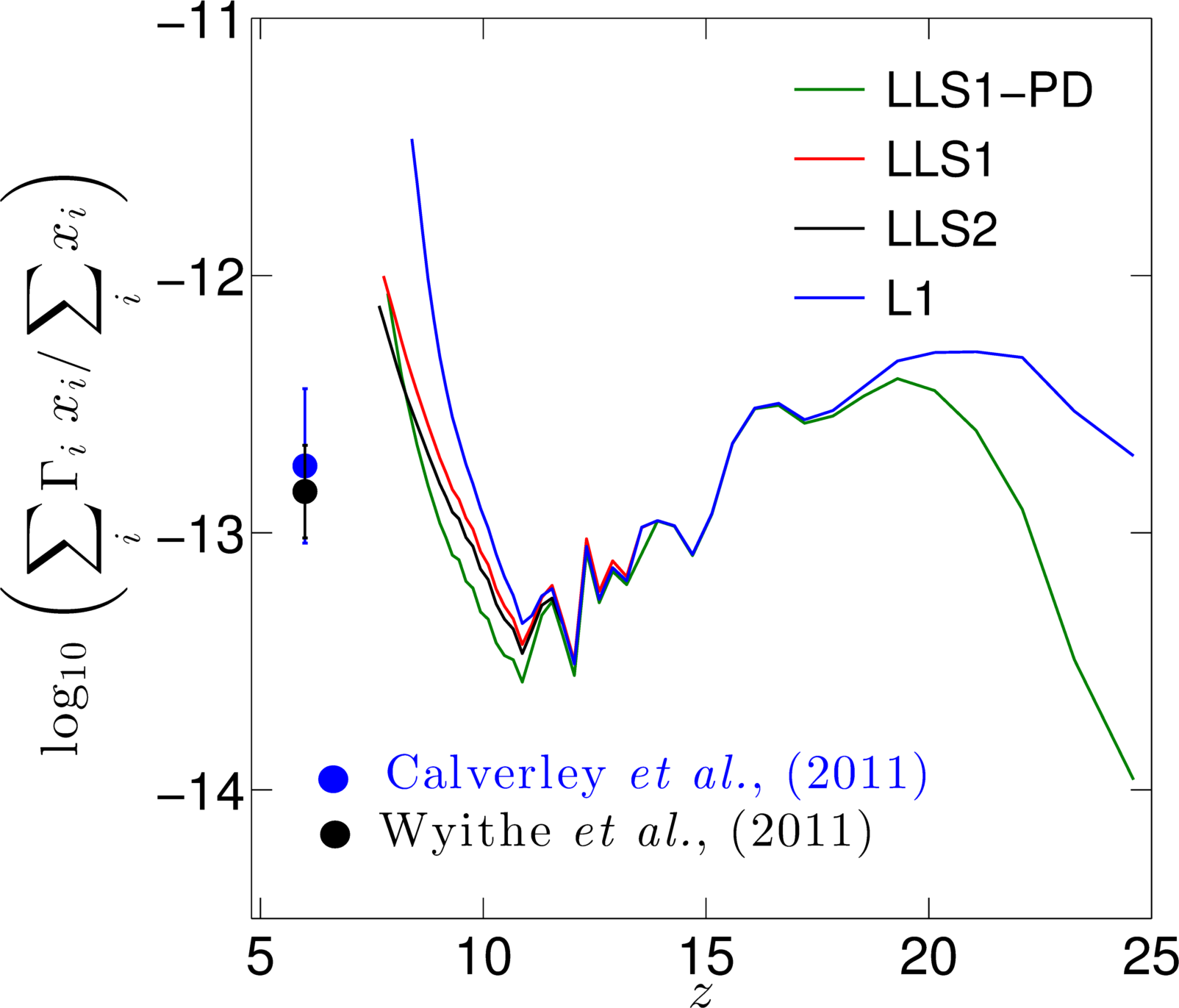}
\caption{The evolution of the ionization fraction-weighted photoionization rates, $\Gamma$ for all models: L1 (blue), LLS1 (red), LLS2 (black) and LLS1-PD (green). The filled circles are observed UV background measured at $z \sim 6$ as $-12.84 \pm 0.18$ \citep{Calverley-2011} (blue) and $-12.74\pm 0.3$ \citep{Wyithe-2011} (black). Both measurements are of 1-$\sigma$ significance.}
\label{fig:ion}
\end{figure}

\subsection {Photoionization rates}
A noteworthy effect of the presence of the spatial variation of the LLS distribution, as shown in Figure \ref{fig:ion}, is that it reduces the mean photoionization rates in the ionized IGM during the peak and late reionization. This can potentially help alleviate the tension that sometimes exist between the measured relatively low values of the photoionization rate (implying a `photon-starved' reionization) and the higher values often found in numerical simulations, which are required to complete reionization by $z\sim6$ and reproduce the measured electron-scattering optical depth. The lower values of the ultraviolet background as measured by \cite{Calverley-2011} and \cite{ Wyithe-2011} at $z\sim 6$, shown in Figure \ref{fig:ion}, are $\Gamma = -12.84 \pm 0.18$ and $-12.74\pm 0.3$ respectively.

The presence of LLS systems increases the opacity in the dense regions and fewer photons can escape into the IGM, which notably reduces the mean photoionization rates regardless of the model details (see Figure~\ref{fig:ion}). This reduction rises towards overlap and reaches a factor of $\sim3$. Interestingly, the position dependent LLS distribution further reduces the mean photoionization rates throughout most of the evolution, roughly doubling the effect compared to the uniform distribution cases (which are very similar to each other). However, the mean photoionization rate for the LLS1-PD case rises fast towards overlap and becomes roughly the same as the uniform LLS cases, while remaining much lower than in the L1 case.     

Although the rise in the photoionization rate overshoots the observed values, the introduction of LLS clearly alleviates the discrepancy. A better match can likely be achieved by lowering the source efficiencies.

The spatial variation of the LLS distribution results in non-trivial changes in the reionization morphology and observable signatures.
Even for a very similar reionization history the ionized patches become smaller and more fragmented, reducing slightly the 21-cm fluctuations, particularly at larger scales, and shifting the peak value to higher frequencies. The most interesting effect of the LLSs is that they reduce the mean photoionization rate in the IGM, which in effect is further enhanced by the spatial variation in the LLS distribution.

%-----------------------------------------
% Section 7 - CONCLUSIONS
%-----------------------------------------

\section{Conclusions} \label {sec-7}

We have presented the results of the first large-scale (114 \mpch) numerical simulations of the epoch of reionization with sources of minimum halo mass $10^8$~M$_\odot$ in the presence of LLSs. The effects of LLS are implemented in C$^2$-Ray using values of the mfp due to LLS extrapolated to the redshift range of reionization from lower redshift observational data, namely, \citet{Songaila-2010aa} and \citet{McQuinn-2011aa}. Instead of imposing a horizon for the distance ionizing photons can travel, the effect of LLS is distributed over the entire IGM, homogeneously or position dependently, in the latter case based on the halo population.

We have analyzed the simulation results with different techniques to explore the underlying physics defining the size distributions, morphologies, and growth rate of the ionized regions in the presence of LLSs and to establish the efficacy of the observable parameters such as the brightness temperature fluctuations and 21-cm power spectra. Our main conclusions are as following,

\begin{itemize}

\item [(i)] We note that by introducing the dampening effects of the LLSs on the ionizing photons, the ionization process is delayed by $\Delta z \sim 0.8$ (for 99\% ionization) for both the LLS models as compared with the fiducial non-LLS model.

\item[(ii)] The photon statistics analysis shows that by the time reionization is complete, the LLS cases have used around 2.5 ionizing photons per baryon, whereas the case without LLS used 1.5. The reduction in the mfp due to LLS thus requires an additional photon per baryon to reionize the Universe. 

\item [(iii)] The topological differences between the large \hii regions in the three models are visible in the simulation data at $\langle x_m\rangle = 0.5$ and indicate a slower merging of ionized regions in the presence of LLS. The friend-of-friend analysis of sizes of \hii regions shows for all cases the emergence of two distinct populations of \hii regions around $\langle x_m\rangle = 0.1$. However, the cases with LLS clearly retain more small \hii regions. The results of the spherical averaging technique for characterizing the sizes of ionized bubbles also show that around $\langle x_m\rangle=0.5$-0.7, the radii of the \hii regions for the LLS cases are smaller by a factor of $\sim$1.2.

\item [(iv)] We also note that while the sizes and mergers of the \hii regions are impeded by the presence of LLSs, the ``freeze out'' of the size as reported in \citet{Sobacchi-2014} is not observed. Both the FoF and spherical averaging methods for characterizing bubble sizes show that the largest \hii regions converge to the entire simulation box towards the end of reionization for both LLS models.

\item [(v)] The 21-cm brightness temperature power spectra show a factor of two less power in fluctuations at large scales ($k < 0.1$ \hmpc) for the LLS cases.

\item [(vi)] The peak value of the RMS of the brightness temperature fluctuations over observable scales (angular scales larger than $3^\prime$ and frequency intervals larger than 0.2~MHz), shows only a decrease of about 9\% when LLS are present.

\item [(vii)] The position dependent LLS results are similar to the homogeneous ones when one considers global quantities. However, the morphologies of ionized regions are affected both in overall shape and due to the presence of neutral islands. Also the mean photoionization rate in the ionized IGM is further reduced in the position dependent LLS case. We conclude that position dependent effects are likely to be important when considering LLS.

\end{itemize}

The method used to implement the LLS in this paper is an improvement over the more usual approach of imposing a hard horizon for ionizing photons. However, it remains ad hoc and depends on the assumed mfp evolution. Our method is useful for ensuring that the mfp in simulations remains bounded to some (redshfit dependent) value but does not capture all of the effects of the sinks of reionization. Major progress in this area will come from high resolution cosmological hydrodynamic radiative transfer simulations and, for the regimes for which these are not feasible, sub-grid models developed based on these. As reionization is a process taking place on the largest cosmological scales but is determined by both sources and sinks on the smallest, galactic scales, it will remain a challenging phenomenon to simulate for the many years to come. 

%-----------------------------------------
% Section 7 - ACKNOWLEDGEMENTS
%-----------------------------------------

\section{Acknowledgements}

This work was supported by the Science and Technology Facilities Council [grant numbers ST/F002858/1 and ST/I000976/1]. HS would like to acknowledge the University of Sussex, School of Mathematical and Physical Sciences fellowship for supporting this work. HS would also like to thank Yi Mao for valuable discussions and the use of the MMRRM code. In addition, HS is grateful to the support from the staff at the supercomputing centers at, University of Portsmouth (Sciama), University of Sussex, Texas Advanced Computing Center, and National Energy Research Supercomputing Center. Finally, HS would also like to express gratitude to Horst Simon at LBNL for his continuing encouragement and support. GM is supported in part by Swedish Research Council grant 2012-4144. We also acknowledge the use of Swedish National Infrastructure for Computing (SNIC) resources at HPC2N (Ume\aa, Sweden). PRS was supported in part by U.S. NSF grants AST-0708176 and AST-1009799, NASA grants NNX07AH09G and NNX11AE09G, and NASA/JPL grant RSA Nos. 1492788 and 1515294, with supercomputer resources from NSF XSEDE grant TG-AST090005 and the TACC at The University of Texas at Austin.

%-----------------------------------------
% Section 8 - REFERENCES
%-----------------------------------------

\bibliographystyle{apj}
\bibliography {LLS_references}

\begin{thebibliography}{}
\expandafter\ifx\csname natexlab\endcsname\relax\def\natexlab#1{#1}\fi

\bibitem[{{Abel} {et~al.}(1999){Abel}, {Norman}, \& {Madau}}]{Abel-1999}
{Abel}, T., {Norman}, M.~L., \& {Madau}, P. 1999, ApJ, 523, 66

\bibitem[{{Alvarez} \& {Abel}(2012)}]{Alvarez-2012}
{Alvarez}, M.~A., \& {Abel}, T. 2012, ApJ, 747, 126

\bibitem[{{Bechtold}(2003)}]{Bechtold-2003}
{Bechtold}, J. 2003, in Galaxies at High Redshift, ed. I.~{P{\'e}rez-Fournon},
  M.~{Balcells}, F.~{Moreno-Insertis}, \& F.~{S{\'a}nchez}, 131--184

\bibitem[{{Calverley} {et~al.}(2011){Calverley}, {Becker}, {Haehnelt}, \&
  {Bolton}}]{Calverley-2011}
{Calverley}, A.~P., {Becker}, G.~D., {Haehnelt}, M.~G., \& {Bolton}, J.~S.
  2011, MNRAS, 412, 2543

\bibitem[{{Choudhury}(2009)}]{Choudhury-2009}
{Choudhury}, T.~R. 2009, Current Science, 97, 841

\bibitem[{{Ciardi}(2006)}]{Ciardi-2006}
{Ciardi}, B. 2006, in IAU Joint Discussion, Vol.~12, IAU Joint Discussion

\bibitem[{Crocce {et~al.}(2006)Crocce, Pueblas, \& Scoccimarro}]{Crocce-2006aa}
Crocce, M., Pueblas, S., \& Scoccimarro, R. 2006, MNRAS, 373, 369

\bibitem[{{Emberson} {et~al.}(2013){Emberson}, {Thomas}, \&
  {Alvarez}}]{Emberson-2013}
{Emberson}, J.~D., {Thomas}, R.~M., \& {Alvarez}, M.~A. 2013, ApJ, 763, 146

\bibitem[{{Erkal}(2014)}]{Erkal-2014}
{Erkal}, D. 2014, ArXiv e-prints

\bibitem[{{Field}(1959)}]{Field-1959}
{Field}, G.~B. 1959, ApJ, 129, 536

\bibitem[{{Friedrich} {et~al.}(2011){Friedrich}, {Mellema}, {Alvarez},
  {Shapiro}, \& {Iliev}}]{Friedrich:2011}
{Friedrich}, M.~M., {Mellema}, G., {Alvarez}, M.~A., {Shapiro}, P.~R., \&
  {Iliev}, I.~T. 2011, MNRAS, 413, 1353

\bibitem[{Furlanetto {et~al.}(2006)Furlanetto, Oh, \& Briggs}]{Furlanetto-2006}
Furlanetto, S., Oh, S.~P., \& Briggs, F. 2006, Phys.Rept., 433, 181

\bibitem[{{Greig} \& {Mesinger}(2015)}]{Greig-2015}
{Greig}, B., \& {Mesinger}, A. 2015, MNRAS, 449, 4246

\bibitem[{{Harnois-D{\'e}raps} {et~al.}(2013){Harnois-D{\'e}raps}, {Pen},
  {Iliev}, {Merz}, {Emberson}, \& {Desjacques}}]{Deraps-2013}
{Harnois-D{\'e}raps}, J., {Pen}, U.-L., {Iliev}, I.~T., {et~al.} 2013, MNRAS,
  436, 540

\bibitem[{{Hinshaw} {et~al.}(2013){Hinshaw}, {Larson}, {Komatsu}, {Spergel},
  {Bennett}, {Dunkley}, {Nolta}, {Halpern}, {Hill}, {Odegard}, {Page}, {Smith},
  {Weiland}, {Gold}, {Jarosik}, {Kogut}, {Limon}, {Meyer}, {Tucker}, {Wollack},
  \& {Wright}}]{Hinshaw-2013}
{Hinshaw}, G., {Larson}, D., {Komatsu}, E., {et~al.} 2013, ApJS, 208, 19

\bibitem[{{Iliev} {et~al.}(2014){Iliev}, {Mellema}, {Ahn}, {Shapiro}, {Mao}, \&
  {Pen}}]{Iliev-2014}
{Iliev}, I.~T., {Mellema}, G., {Ahn}, K., {et~al.} 2014, MNRAS, 439, 725

\bibitem[{{Iliev} {et~al.}(2008){Iliev}, {Mellema}, {Pen}, {Bond}, \&
  {Shapiro}}]{Iliev-2008}
{Iliev}, I.~T., {Mellema}, G., {Pen}, U.-L., {Bond}, J.~R., \& {Shapiro}, P.~R.
  2008, MNRAS., 384, 863

\bibitem[{{Iliev} {et~al.}(2006){Iliev}, {Mellema}, {Pen}, {Merz}, {Shapiro},
  \& {Alvarez}}]{Iliev-2006}
{Iliev}, I.~T., {Mellema}, G., {Pen}, U.-L., {et~al.} 2006, MNRAS, 369, 1625

\bibitem[{Iliev {et~al.}(2006)Iliev, Mellema, Pen, Merz, Shapiro, \&
  Alvarez}]{Iliev-2006aa}
Iliev, I.~T., Mellema, G., Pen, U.-L., {et~al.} 2006, Mon.Not.Roy.Astron.Soc.,
  369, 1625

\bibitem[{{Iliev} {et~al.}(2007){Iliev}, {Mellema}, {Shapiro}, \&
  {Pen}}]{Iliev-2007}
{Iliev}, I.~T., {Mellema}, G., {Shapiro}, P.~R., \& {Pen}, U.-L. 2007,
  Mon.Not.Roy.Astron.Soc., 376, 534

\bibitem[{{Iliev} {et~al.}(2012){Iliev}, {Mellema}, {Shapiro}, {Pen}, {Mao},
  {Koda}, \& {Ahn}}]{Iliev-2012}
{Iliev}, I.~T., {Mellema}, G., {Shapiro}, P.~R., {et~al.} 2012,
  Mon.Not.Roy.Astron.Soc., 423, 2222

\bibitem[{{Iliev} {et~al.}(2005{\natexlab{a}}){Iliev}, {Scannapieco}, \&
  {Shapiro}}]{Iliev-2005a}
{Iliev}, I.~T., {Scannapieco}, E., \& {Shapiro}, P.~R. 2005{\natexlab{a}}, ApJ,
  624, 491

\bibitem[{{Iliev} {et~al.}(2005{\natexlab{b}}){Iliev}, {Shapiro}, \&
  {Raga}}]{Shapiro-2005}
{Iliev}, I.~T., {Shapiro}, P.~R., \& {Raga}, A.~C. 2005{\natexlab{b}}, MNRAS,
  361, 405

\bibitem[{{Iliev} {et~al.}(2005{\natexlab{c}}){Iliev}, {Shapiro}, \&
  {Raga}}]{Iliev-2005}
---. 2005{\natexlab{c}}, MNRAS, 361, 405

\bibitem[{{Kaurov} \& {Gnedin}(2013)}]{Kaurov-2013}
{Kaurov}, A.~A., \& {Gnedin}, N.~Y. 2013, ApJ, 771, 35

\bibitem[{{Kohler} \& {Gnedin}(2007)}]{Kohler-2007}
{Kohler}, K., \& {Gnedin}, N.~Y. 2007, ApJ, 655, 685

\bibitem[{Lewis {et~al.}(2000)Lewis, Challinor, \& Lasenby}]{Lewis-1999}
Lewis, A., Challinor, A., \& Lasenby, A. 2000, Astrophys. J., 538, 473

\bibitem[{{Mao} {et~al.}(2012){Mao}, {Shapiro}, {Mellema}, {Iliev}, {Koda}, \&
  {Ahn}}]{Mao-2012}
{Mao}, Y., {Shapiro}, P.~R., {Mellema}, G., {et~al.} 2012, MNRAS, 422, 926

\bibitem[{{McQuinn} {et~al.}(2007){McQuinn}, {Lidz}, {Zahn}, {Dutta},
  {Hernquist}, \& {Zaldarriaga}}]{McQuinn-2007}
{McQuinn}, M., {Lidz}, A., {Zahn}, O., {et~al.} 2007, Mon.Not.Roy.Astron.Soc.,
  377, 1043

\bibitem[{McQuinn {et~al.}(2011)McQuinn, Oh, \&
  Faucher-Giguere}]{McQuinn-2011aa}
McQuinn, M., Oh, S.~P., \& Faucher-Giguere, C.-A. 2011, Astrophys.J.743:82,
  2011

\bibitem[{Mellema {et~al.}(2006)Mellema, Iliev, Alvarez, \&
  Shapiro}]{Mellema-2006a}
Mellema, G., Iliev, I.~T., Alvarez, M.~A., \& Shapiro, P.~R. 2006, New
  Astronomy, 11, 374

\bibitem[{{Mesinger} \& {Furlanetto}(2007)}]{Mesinger-2007}
{Mesinger}, A., \& {Furlanetto}, S. 2007, ApJ, 669, 663

\bibitem[{{Mesinger} {et~al.}(2011){Mesinger}, {Furlanetto}, \&
  {Cen}}]{Mesinger-2010}
{Mesinger}, A., {Furlanetto}, S., \& {Cen}, R. 2011, MNRAS, 411, 955

\bibitem[{{Miralda-Escud{\'e}}(2003)}]{Miralda-Escude-2003}
{Miralda-Escud{\'e}}, J. 2003, ApJ, 597, 66

\bibitem[{{Osterbrock}(1989)}]{Osterbock-1989}
{Osterbrock}, D.~E. 1989, {Astrophysics of gaseous nebulae and active galactic
  nuclei} (University Science Books)

\bibitem[{{Patil} {et~al.}(2014){Patil}, {Zaroubi}, {Chapman}, {Jeli{\'c}},
  {Harker}, {Abdalla}, {Asad}, {Bernardi}, {Brentjens}, {de Bruyn}, {Bus},
  {Ciardi}, {Daiboo}, {Fernandez}, {Ghosh}, {Jensen}, {Kazemi}, {Koopmans},
  {Labropoulos}, {Mevius}, {Martinez}, {Mellema}, {Offringa}, {Pandey},
  {Schaye}, {Thomas}, {Vedantham}, {Veligatla}, {Wijnholds}, \&
  {Yatawatta}}]{Patil-2014}
{Patil}, A.~H., {Zaroubi}, S., {Chapman}, E., {et~al.} 2014, MNRAS, 443, 1113

\bibitem[{{Petitjean} {et~al.}(1993){Petitjean}, {Webb}, {Rauch}, {Carswell},
  \& {Lanzetta}}]{Petitjean:1993}
{Petitjean}, P., {Webb}, J.~K., {Rauch}, M., {Carswell}, R.~F., \& {Lanzetta},
  K. 1993, MNRAS, 262, 499

\bibitem[{{Planck Collaboration} {et~al.}(2015){Planck Collaboration}, {Ade},
  {Aghanim}, {Arnaud}, {Ashdown}, {Aumont}, {Baccigalupi}, {Banday},
  {Barreiro}, {Bartlett}, \& et~al.}]{Ade-2015-b}
{Planck Collaboration}, {Ade}, P.~A.~R., {Aghanim}, N., {et~al.} 2015, ArXiv
  e-prints

\bibitem[{{Press} {et~al.}(1992){Press}, {Teukolsky}, {Vetterling}, \&
  {Flannery}}]{Press-1992}
{Press}, W.~H., {Teukolsky}, S.~A., {Vetterling}, W.~T., \& {Flannery}, B.~P.
  1992, {Numerical recipes in FORTRAN. The art of scientific computing} (Press
  Syndicate of the University of Cambridge)

\bibitem[{{Prochaska} {et~al.}(2010){Prochaska}, {O'Meara}, \&
  {Worseck}}]{Prochaska-2010}
{Prochaska}, J.~X., {O'Meara}, J.~M., \& {Worseck}, G. 2010, ApJ, 718, 392

\bibitem[{{Ribaudo} {et~al.}(2011){Ribaudo}, {Lehner}, \&
  {Howk}}]{Ribaudo-2011}
{Ribaudo}, J., {Lehner}, N., \& {Howk}, J.~C. 2011, ApJ, 736, 42

\bibitem[{{Santos} {et~al.}(2010){Santos}, {Ferramacho}, {Silva}, {Amblard}, \&
  {Cooray}}]{Santos-2010}
{Santos}, M.~G., {Ferramacho}, L., {Silva}, M.~B., {Amblard}, A., \& {Cooray},
  A. 2010, MNRAS, 406, 2421

\bibitem[{{Seljak} \& {Zaldarriaga}(1996)}]{Seljak-1996}
{Seljak}, U., \& {Zaldarriaga}, M. 1996, ApJ, 469, 437

\bibitem[{{Shapiro} {et~al.}(2006){Shapiro}, {Iliev}, {Alvarez}, \&
  {Scannapieco}}]{Shapiro-2006}
{Shapiro}, P.~R., {Iliev}, I.~T., {Alvarez}, M.~A., \& {Scannapieco}, E. 2006,
  ApJ, 648, 922

\bibitem[{{Shapiro} {et~al.}(2004){Shapiro}, {Iliev}, \& {Raga}}]{Shapiro-2004}
{Shapiro}, P.~R., {Iliev}, I.~T., \& {Raga}, A.~C. 2004, MNRAS, 348, 753

\bibitem[{{Sobacchi} \& {Mesinger}(2014)}]{Sobacchi-2014}
{Sobacchi}, E., \& {Mesinger}, A. 2014, MNRAS, 440, 1662

\bibitem[{Songaila \& Cowie(2010)}]{Songaila-2010aa}
Songaila, A., \& Cowie, L.~L. 2010, ApJ,, 721, 1448

\bibitem[{Tytler(1982)}]{Tytler-1982}
Tytler, D. 1982, Nature, 298, 427

\bibitem[{{Watson} {et~al.}(2013){Watson}, {Iliev}, {D'Aloisio}, {Knebe},
  {Shapiro}, \& {Yepes}}]{Watson-2013}
{Watson}, W.~A., {Iliev}, I.~T., {D'Aloisio}, A., {et~al.} 2013, MNRAS, 433,
  1230

\bibitem[{{Worseck} {et~al.}(2014){Worseck}, {Prochaska}, {O'Meara}, {Becker},
  {Ellison}, {Lopez}, {Meiksin}, {M{\'e}nard}, {Murphy}, \&
  {Fumagalli}}]{Worseck-2014}
{Worseck}, G., {Prochaska}, J.~X., {O'Meara}, J.~M., {et~al.} 2014, MNRAS, 445,
  1745

\bibitem[{{Wyithe} \& {Bolton}(2011)}]{Wyithe-2011}
{Wyithe}, J.~S.~B., \& {Bolton}, J.~S. 2011, MNRAS, 412, 1926

\bibitem[{{Zahn} {et~al.}(2007){Zahn}, {Lidz}, {McQuinn}, {Dutta}, {Hernquist},
  {Zaldarriaga}, \& {Furlanetto}}]{Zahn-2007}
{Zahn}, O., {Lidz}, A., {McQuinn}, M., {et~al.} 2007, ApJ, 654, 12

\end{thebibliography}

\end{document}